\documentclass{article}
\usepackage{Odyssey2020}
\usepackage{amsmath,graphicx}

\usepackage{multirow}
\usepackage{graphicx}
\usepackage{amssymb, amsmath, bm, mathtools,array}
\usepackage{textcomp}
\usepackage{hyperref}
\usepackage{verbatim,lipsum}
\usepackage{booktabs}
\usepackage{tcolorbox}
\usepackage{lscape}
\usepackage{subcaption}
\usepackage{multirow}
\usepackage{colortbl}
\usepackage{color}

\RequirePackage[normalem]{ulem} 
\RequirePackage{color}\definecolor{BLUE}{rgb}{0,0.20,0.75} 
\RequirePackage{color}\definecolor{BROWN}{RGB}{60,128,49} 

\definecolor{mygreen}{RGB}{1, 183, 183}
\definecolor{mygrey}{RGB}{127, 127, 127}

\newcommand{\bs}[1]{\boldsymbol{#1}}
\newcommand{\BNIshort}{\texttt{LLGF}}
\newcommand{\BNIIshort}{\texttt{LGF}}
\newcommand{\BNIIIshort}{\texttt{GF}}

\newcommand{\BNI}{\texttt{LLGF}}
\newcommand{\BNII}{\texttt{LGF}}
\newcommand{\BNIII}{\texttt{GF}}

\title{Investigating self-supervised front ends for speech spoofing countermeasures}
\name{Xin Wang\thanks{This study is supported by JST CREST Grants (JPMJCR18A6 and JPMJCR20D3), MEXT KAKENHI Grants (21K17775, 21H04906, 18H04112, 16H06302), and Google AI for Japan program.}, Junichi Yamagishi}
\address{National Institute of Informatics, Japan}
\begin{document}
\ninept
\maketitle
\begin{abstract}
Self-supervised speech model is a rapid progressing research topic, and many pre-trained models have been released and used in various down stream tasks. For speech anti-spoofing, most countermeasures (CMs) use signal processing algorithms to extract acoustic features for classification. In this study, we use pre-trained self-supervised speech models as the front end of spoofing CMs. We investigated different back end architectures to be combined with the self-supervised front end, the effectiveness of fine-tuning the front end, and the performance of using different pre-trained self-supervised models. Our findings showed that, when a good pre-trained front end was fine-tuned with either a shallow or a deep neural network-based back end on the ASVspoof 2019 logical access (LA) training set, the resulting CM not only achieved a low EER score on the 2019 LA test set but also significantly outperformed the baseline on the ASVspoof 2015, 2021 LA, and 2021 deepfake test sets. A sub-band analysis further demonstrated that the CM mainly used the information in a specific frequency band to discriminate the bona fide and spoofed trials across the test sets.
\end{abstract}
%

%
\section{Introduction}
\label{sec:intro}
Advanced voice conversion and text-to-speech technologies can be misused to attack automatic speaker verification (ASV) systems \cite{evans2013spoofing} or fool humans. 
Protecting ASV systems and human users from the threat of spoofed speech calls for reliable automatic spoofing countermeasures (CMs). 

Most spoofing CMs consist of a front end and a back end. The front end extracts $N$ frames of acoustic features $\bs{a}_{1:N}$ from an input speech trial $\bs{x}_{1:T}$ of length $T$, and the back end converts $\bs{a}_{1:N}$ into a score $s\in\mathbb{R}$ that indicates how likely the input trial is spoofed or bona fide (i.e., real human speech). 
Most conventional front ends rely on digital signal processing (DSP) algorithms to extract spectra, phase, or other acoustic features \cite{kamble_sailor_patil_li_2020}. The most widely used acoustic features include linear frequency cepstral coefficients (LFCC) \cite{davis1980comparison} and Constant-Q cepstrum coefficients (CQCC) \cite{Todisco2017}.

While the hard-wired DSP front end performed well on many benchmark databases \cite{Todisco2017, sahidullah2015comparison}, the research community has proposed many methods to make the front end trainable. The motivation is to encourage the front end to extract more discriminative acoustic features for the anti-spoofing task. One thread of work tries to integrate the DSP with deep neural networks (DNNs) by, for example, replacing the linear scale filter bank with a hidden layer from a pre-trained DNN \cite{yu2017dnn}.  
A similar method uses a DNN to predict the center frequency of each filter in the filter bank \cite{fu2021fastaudio}.
Another example is the trainable windowed-sinc filter proposed in SincNet \cite{ravanelli2018speaker}, which has been used in one CM \cite{tak2020end}. These DNN-based front ends are trained in a supervised manner using an anti-spoofing database.

When using either a DSP or DNN-based front end, a CM well trained on a closed-set benchmark database can significantly degrade when faced unknown spoofing attacks or bona fide trials from mismatched domains \cite{paul2017generalization,das2020assessing,yamagishi21_asvspoof}. Designing a DSP-based CM front end robust to mismatched domains is an ongoing topic \cite{das2019long}. On the other hand, training a robust supervised DNN front end requires a sufficient amount of bona fide and spoofed speech data. However, generating spoofed trials is laborious and technically demanding. 

These difficulties motivate us to use a self-supervised speech model as the CM front end.
The idea is to use a DNN to extract the acoustic features $\bs{a}_{1:N}$, but the DNN is trained in a self-supervised manner. Such a DNN requires no spoofed trials and can be trained on any speech database.  With a great variety of training data, the self-supervised model may extract acoustic features robust to the unknown domains for the CM task. 
Although training a good self-supervised speech model is costly,  many pre-trained self-supervised models are available and can be used off the shelf.

This study investigates the effectiveness of using the pre-trained self-supervised model as the CM front end. Specifically,  
\begin{enumerate}
\setlength\itemsep{-0.2em}
\item What kind of back end architecture is suitable for a self-supervised front end?
\item Should the pre-trained self-supervised front end be fine-tuned for the anti-spoofing task?
\item Among the many publicly available pre-trained self-supervised models, which one is best for anti-spoofing?
\end{enumerate}
Our experiments were conducted using the ASVspoof 2019 logical access (LA) training set \cite{Todisco2019} and multiple test sets from the ASVspoof 2015, 2019, and 2021 challenges \cite{yamagishi21_asvspoof, wu2015asvspoof}.
The results suggest that the back end needs to be deep when the pre-trained front end is not fine-tuned for the anti-spoofing task. 
However, if the front end can be fine-tuned with the rest of the CM, a simple back end with just an average temporal pooling and linear layer is sufficient. The resulting CM not only performed equally as well as a strong baseline CM on the LA 2019 test set but also significantly reduced the equal error rate (EER) on all the other test sets. 
As for the last question, a model pre-trained on diverse speech corpora is recommended. 
These experiments and findings hence differentiate this study from other works that used only one self-supervised model on one test set \cite{jiang20b_interspeech, xie21_interspeech}.  Furthermore, a sub-band analysis on the CMs was conducted, and the results showed interesting differences between the CMs using a self-supervised front end and the baseline using LFCC. 
Code will be released upon publication\footnote{https://github.com/nii-yamagishilab/project-NN-Pytorch-scripts}.

In the rest of this paper, Section 2 briefly describes the self-supervised models used in this study. Section 3 and 4 details the experiments and sub-band analysis, respectively. Section 5 draws a conclusion.

\begin{figure}[t]
\centering
\includegraphics[trim=0 220 0 90, clip, width=0.95\columnwidth]{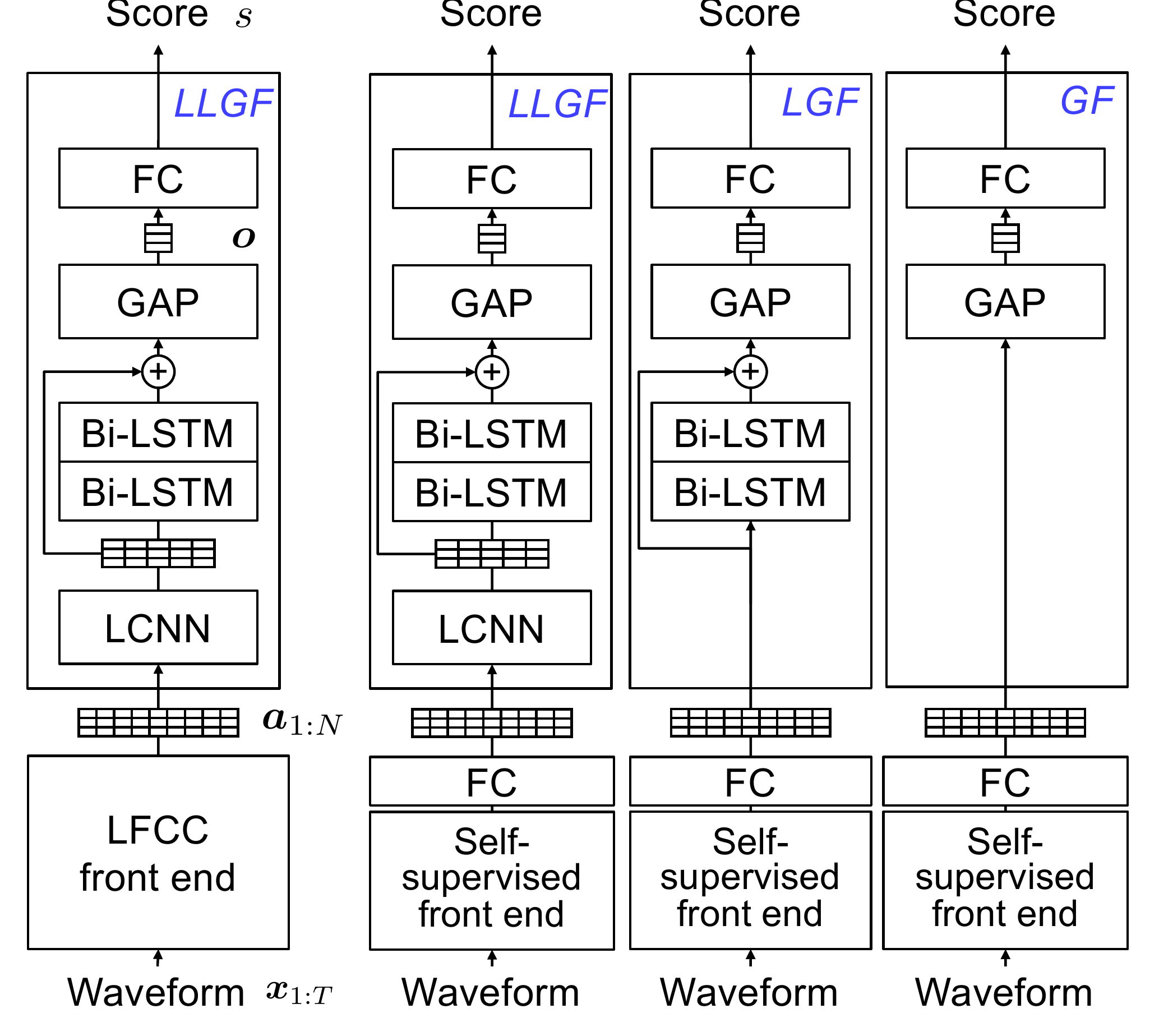}
\vspace{-3mm}
\caption{Baseline and CMs using a self-supervised front end. GAP and FC denote global average pooling and fully connected layers, respectively. \BNI, \BNII, and \BNIII\ denote the three back end types.}
\vspace{-5mm}
\label{fig:idea}
\end{figure}

\section{Methods}
\label{sec:ssl}
\subsection{Self-supervised speech models}
Among many self-supervised speech models currently in use, this study focuses on Wav2vec 2.0 \cite{NEURIPS2020_92d1e1eb} and HuBERT \cite{hsu2021hubert}. 
Wav2vec 2.0 consists of a convolution neural network (CNN) and a Transformer \cite{vaswani2017attention}. The former extracts a sequence of feature vectors $\bs{z}_{1:N}$ from the input waveform $\bs{x}_{1:T}$, and the latter transforms $\bs{z}_{1:N}$ into the output $\bs{a}_{1:N}$ that \emph{captures the information from the entire sequence} \cite{NEURIPS2020_92d1e1eb}. Note that the ratio between the $N$ and $T$ is decided by the CNN stride and is equal to $N/T = 1/320$ in a default setting. 
During training, the model quantizes the latent $\bs{z}_{1:N}$ into $\bs{q}_{1:N}$, masks part of $\bs{z}_{1:N}$, and computes a new sequence $\bs{c}_{1:N}$ from the Transformer given the partially masked $\bs{z}_{1:N}$. The loss measures how well the model identifies each $q_{n}$ among multiple distractors given $\bs{c}_n$.  HuBERT is similar to Wav2vec 2.0 but uses a different training criterion and procedure. 

\subsection{CM with a self-supervised front end}
By feeding the output $\bs{a}_{1:N}$ from the self-supervised model to the back end,  the CM can obtain a score $s\in\mathbb{R}$ for the input waveform. 
However, to find the best configuration for such a CM, we need to consider the following factors.

\subsubsection{Back end architecture}
Some studies have suggested that a shallow network is sufficient as the back end for down stream tasks \cite{NEURIPS2020_92d1e1eb, yang21c_interspeech}, 
but such a claim has to be verified empirically for the anti-spoofing task. 
We compared three types of back ends, as shown on the right side of Figure~\ref{fig:idea}. 
The first one is taken from a standard baseline CM \cite{wang2021comparative, yamagishi21_asvspoof}: a light convolution neural network (LCNN) followed by two bi-directional recurrent layers using long short-term memory (LSTM) units, a global average pooling layer, and a fully connected (FC) output layer. It is referred to as \BNI. Note that this back end has achieved good performance on the ASVspoof 2019 LA database \cite{wang2021comparative}.  
As for the other two types, \BNII\ removes the LCNN part, and \BNIII\ further removes the LSTM layers. 

Note that an FC layer is inserted between the back end and the self-supervised front end. This reduces the dimensions of the self-supervised model's output and is jointly trained with the back end. 

\subsubsection{Fine-tune or freeze pre-trained self-supervised front end}
In some applications, a pre-trained self-supervised model can be used without fine-tuning \cite{yang21c_interspeech}. However, some studies have found fine-tuning beneficial \cite{NEURIPS2020_92d1e1eb}. We test both strategies in this study. For the fine-tuning case, we simply initialize the self-supervised front end using a pre-trained model and further train the front end with the rest of the CM on the CM training set. The training recipe is detailed in the next section.

\subsubsection{Different pre-trained self-supervised front ends}
Finally, there are many pre-trained self-supervised models released online. Some have been trained using speech data from various corpora, while others use only a limited amount of data. 
We compare a few pre-trained models released by the Fairseq project \cite{ott2019fairseq}\footnote{\url{https://github.com/pytorch/fairseq/blob/main/examples/wav2vec}, \url{https://github.com/pytorch/fairseq/tree/main/examples/HuBERT}} for the CM. Their details are listed in Table~\ref{tab:ssl}.

\begin{table}[t]
\caption{Overview of self-supervised models used in this study.}
\vspace{-5mm}
\begin{center}
\resizebox{\columnwidth}{!}{
\setlength{\tabcolsep}{2pt}
\begin{tabular}{llp{3.5cm}rr}
\toprule
    ID      & Model type  & Training data & \#.para & Out dim. \\
\midrule
\texttt{W2V-XLSR} & Wav2vec (xlsr) & LibriSpeech \cite{panayotov2015librispeech}, CommonVoice \cite{ardila-etal-2020-common}, BABEL \cite{harper2011iarpa} & 317 m & 1024 \\
\texttt{W2V-Large2} & Wav2vec (w2v\_large) & CommonVoice, Switchboard \cite{godfrey1992switchboard}, Libri-Light \cite{kahn2020libri}, Fisher \cite{cieri-etal-2004-fisher} & 317 m & 1024 \\
\texttt{W2V-Large1} & Wav2vec (w2v\_vox\_new) & Libri-Light  & 317 m & 1024 \\
\texttt{W2V-Small} & Wav2vec (w2v\_small) & Librispeech & 95 m & 768 \\
\texttt{HuBERT-XL} & HuBERT (extra\_large) & Libri-Light &  964 m & 1280 \\
\bottomrule
\end{tabular}
}
\end{center}
\vspace{-5mm}
\label{tab:ssl}
\end{table}

\begin{table*}[t]
\caption{EERs (\%) on different test sets. All the models were trained using the ASVspoof 2019 LA training set. A darker cell color indicates a higher EER value. Different back end types are illustrated in Figure~\ref{fig:idea}. For visualization, the results of the three training-evaluation rounds are sorted in accordance with EER on the LA 2019 test set from low (I) to high (III). Results of the model using \texttt{W2V-XLSR} and \BNI\ are copied to the 2nd sub-table. }
\vspace{-5mm}
\begin{center}
\resizebox{\textwidth}{!}
{
\setlength{\tabcolsep}{4pt}
\begin{tabular}{l|ccc|ccc|ccc|ccc|ccc|ccc|ccc}
\toprule
Front end  &    \multicolumn{3}{c|}{LFCC} & \multicolumn{9}{c|}{\texttt{W2V-XLSR}, fixed} & \multicolumn{9}{c|}{\texttt{W2V-XLSR}, fine-tuned} \\
 \midrule
Back end &  \multicolumn{3}{c|}{\BNI} & \multicolumn{3}{c|}{\BNI} & \multicolumn{3}{c|}{\BNII} & \multicolumn{3}{c|}{\BNIII} & \multicolumn{3}{c|}{\BNI} & \multicolumn{3}{c|}{\BNII} & \multicolumn{3}{c|}{\BNIII} \\
 \midrule
              &   I   &  II   &  III  &   I   &  II   &  III  &   I   &  II   &  III  &   I   &  II   &  III  &   I   &  II   &  III  &   I   &  II   &  III  &   I   &  II   &  III \\ 
\midrule
   2019 LA    & \cellcolor[rgb]{0.99, 0.99, 0.99} 2.98 & \cellcolor[rgb]{0.99, 0.99, 0.99} 3.03 & \cellcolor[rgb]{0.98, 0.98, 0.98} 3.33 & \cellcolor[rgb]{0.99, 0.99, 0.99} 1.47 & \cellcolor[rgb]{0.98, 0.98, 0.98} 3.45 & \cellcolor[rgb]{0.98, 0.98, 0.98} 3.77 & \cellcolor[rgb]{0.97, 0.97, 0.97} 6.01 & \cellcolor[rgb]{0.97, 0.97, 0.97} 6.32 & \cellcolor[rgb]{0.96, 0.96, 0.96} 6.95 & \cellcolor[rgb]{0.90, 0.90, 0.90} 15.96 & \cellcolor[rgb]{0.89, 0.89, 0.89} 16.72 & \cellcolor[rgb]{0.89, 0.89, 0.89} 16.98 & \cellcolor[rgb]{0.99, 0.99, 0.99} 2.31 & \cellcolor[rgb]{0.99, 0.99, 0.99} 2.80 & \cellcolor[rgb]{0.99, 0.99, 0.99} 3.08 & \cellcolor[rgb]{0.99, 0.99, 0.99} 1.28 & \cellcolor[rgb]{0.99, 0.99, 0.99} 1.28 & \cellcolor[rgb]{0.99, 0.99, 0.99} 1.50 & \cellcolor[rgb]{0.99, 0.99, 0.99} 1.96 & \cellcolor[rgb]{0.99, 0.99, 0.99} 2.25 & \cellcolor[rgb]{0.99, 0.99, 0.99} 2.27\\ 
   2015 LA    & \cellcolor[rgb]{0.77, 0.77, 0.77} 29.42 & \cellcolor[rgb]{0.78, 0.78, 0.78} 27.98 & \cellcolor[rgb]{0.75, 0.75, 0.75} 31.21 & \cellcolor[rgb]{0.98, 0.98, 0.98} 3.97 & \cellcolor[rgb]{0.96, 0.96, 0.96} 6.78 & \cellcolor[rgb]{0.96, 0.96, 0.96} 8.18 & \cellcolor[rgb]{0.94, 0.94, 0.94} 10.04 & \cellcolor[rgb]{0.94, 0.94, 0.94} 10.95 & \cellcolor[rgb]{0.95, 0.95, 0.95} 9.51 & \cellcolor[rgb]{0.89, 0.89, 0.89} 16.90 & \cellcolor[rgb]{0.88, 0.88, 0.88} 17.55 & \cellcolor[rgb]{0.88, 0.88, 0.88} 17.89 & \cellcolor[rgb]{1.00, 1.00, 1.00} 0.25 & \cellcolor[rgb]{1.00, 1.00, 1.00} 0.41 & \cellcolor[rgb]{1.00, 1.00, 1.00} 0.24 & \cellcolor[rgb]{1.00, 1.00, 1.00} 0.24 & \cellcolor[rgb]{1.00, 1.00, 1.00} 0.19 & \cellcolor[rgb]{1.00, 1.00, 1.00} 0.31 & \cellcolor[rgb]{1.00, 1.00, 1.00} 0.21 & \cellcolor[rgb]{1.00, 1.00, 1.00} 0.17 & \cellcolor[rgb]{1.00, 1.00, 1.00} 0.17\\ 
2021 LA prog. & \cellcolor[rgb]{0.90, 0.90, 0.90} 15.82 & \cellcolor[rgb]{0.90, 0.90, 0.90} 15.81 & \cellcolor[rgb]{0.82, 0.82, 0.82} 24.40 & \cellcolor[rgb]{0.95, 0.95, 0.95} 9.85 & \cellcolor[rgb]{0.88, 0.88, 0.88} 17.29 & \cellcolor[rgb]{0.86, 0.86, 0.86} 20.17 & \cellcolor[rgb]{0.89, 0.89, 0.89} 16.76 & \cellcolor[rgb]{0.92, 0.92, 0.92} 13.77 & \cellcolor[rgb]{0.90, 0.90, 0.90} 15.63 & \cellcolor[rgb]{0.86, 0.86, 0.86} 20.06 & \cellcolor[rgb]{0.86, 0.86, 0.86} 20.88 & \cellcolor[rgb]{0.85, 0.85, 0.85} 21.25 & \cellcolor[rgb]{0.96, 0.96, 0.96} 7.58 & \cellcolor[rgb]{0.97, 0.97, 0.97} 6.38 & \cellcolor[rgb]{0.96, 0.96, 0.96} 6.56 & \cellcolor[rgb]{0.94, 0.94, 0.94} 10.63 & \cellcolor[rgb]{0.95, 0.95, 0.95} 9.19 & \cellcolor[rgb]{0.97, 0.97, 0.97} 6.27 & \cellcolor[rgb]{0.96, 0.96, 0.96} 7.65 & \cellcolor[rgb]{0.96, 0.96, 0.96} 7.16 & \cellcolor[rgb]{0.96, 0.96, 0.96} 7.82\\ 
2021 LA eval. & \cellcolor[rgb]{0.85, 0.85, 0.85} 20.93 & \cellcolor[rgb]{0.86, 0.86, 0.86} 20.38 & \cellcolor[rgb]{0.79, 0.79, 0.79} 27.06 & \cellcolor[rgb]{0.94, 0.94, 0.94} 10.97 & \cellcolor[rgb]{0.87, 0.87, 0.87} 18.91 & \cellcolor[rgb]{0.86, 0.86, 0.86} 20.71 & \cellcolor[rgb]{0.86, 0.86, 0.86} 20.23 & \cellcolor[rgb]{0.90, 0.90, 0.90} 16.02 & \cellcolor[rgb]{0.89, 0.89, 0.89} 16.52 & \cellcolor[rgb]{0.86, 0.86, 0.86} 20.30 & \cellcolor[rgb]{0.85, 0.85, 0.85} 21.16 & \cellcolor[rgb]{0.85, 0.85, 0.85} 21.48 & \cellcolor[rgb]{0.96, 0.96, 0.96} 7.62 & \cellcolor[rgb]{0.96, 0.96, 0.96} 7.26 & \cellcolor[rgb]{0.96, 0.96, 0.96} 7.18 & \cellcolor[rgb]{0.95, 0.95, 0.95} 9.66 & \cellcolor[rgb]{0.96, 0.96, 0.96} 8.11 & \cellcolor[rgb]{0.96, 0.96, 0.96} 6.53 & \cellcolor[rgb]{0.96, 0.96, 0.96} 7.99 & \cellcolor[rgb]{0.96, 0.96, 0.96} 7.42 & \cellcolor[rgb]{0.96, 0.96, 0.96} 7.61\\
2021 DF prog. & \cellcolor[rgb]{0.78, 0.78, 0.78} 28.38 & \cellcolor[rgb]{0.83, 0.83, 0.83} 23.60 & \cellcolor[rgb]{0.75, 0.75, 0.75} 31.12 & \cellcolor[rgb]{0.99, 0.99, 0.99} 2.67 & \cellcolor[rgb]{0.97, 0.97, 0.97} 5.09 & \cellcolor[rgb]{0.96, 0.96, 0.96} 7.02 & \cellcolor[rgb]{0.96, 0.96, 0.96} 6.92 & \cellcolor[rgb]{0.96, 0.96, 0.96} 7.91 & \cellcolor[rgb]{0.96, 0.96, 0.96} 8.39 & \cellcolor[rgb]{0.87, 0.87, 0.87} 19.30 & \cellcolor[rgb]{0.86, 0.86, 0.86} 20.26 & \cellcolor[rgb]{0.86, 0.86, 0.86} 20.63 & \cellcolor[rgb]{0.98, 0.98, 0.98} 4.40 & \cellcolor[rgb]{0.98, 0.98, 0.98} 4.33 & \cellcolor[rgb]{0.98, 0.98, 0.98} 4.14 & \cellcolor[rgb]{0.98, 0.98, 0.98} 3.38 & \cellcolor[rgb]{0.98, 0.98, 0.98} 3.75 & \cellcolor[rgb]{0.98, 0.98, 0.98} 3.55 & \cellcolor[rgb]{0.98, 0.98, 0.98} 3.97 & \cellcolor[rgb]{0.98, 0.98, 0.98} 4.23 &    \cellcolor[rgb]{0.97, 0.97, 0.97} 4.94  \\ 
2021 DF eval. & \cellcolor[rgb]{0.82, 0.82, 0.82} 24.37 & \cellcolor[rgb]{0.83, 0.83, 0.83} 23.05 & \cellcolor[rgb]{0.79, 0.79, 0.79} 27.22 & \cellcolor[rgb]{0.96, 0.96, 0.96} 7.14 & \cellcolor[rgb]{0.95, 0.95, 0.95} 9.94 & \cellcolor[rgb]{0.94, 0.94, 0.94} 11.35 & \cellcolor[rgb]{0.92, 0.92, 0.92} 13.26 & \cellcolor[rgb]{0.92, 0.92, 0.92} 13.23 & \cellcolor[rgb]{0.93, 0.93, 0.93} 12.00 & \cellcolor[rgb]{0.87, 0.87, 0.87} 18.88 & \cellcolor[rgb]{0.87, 0.87, 0.87} 19.48 & \cellcolor[rgb]{0.86, 0.86, 0.86} 19.81 & \cellcolor[rgb]{0.97, 0.97, 0.97} 5.44 & \cellcolor[rgb]{0.96, 0.96, 0.96} 6.68 & \cellcolor[rgb]{0.97, 0.97, 0.97} 6.18 & \cellcolor[rgb]{0.98, 0.98, 0.98} 4.75 & \cellcolor[rgb]{0.97, 0.97, 0.97} 5.23 & \cellcolor[rgb]{0.97, 0.97, 0.97} 4.98 & \cellcolor[rgb]{0.97, 0.97, 0.97} 5.04 & \cellcolor[rgb]{0.97, 0.97, 0.97} 6.10 &   \cellcolor[rgb]{0.97, 0.97, 0.97} 5.88   \\ 
\bottomrule
\multicolumn{4}{c}{} & \multicolumn{3}{c}{($\Downarrow$ results are copied)} & \multicolumn{12}{c}{} & $\phantom{00.00}$ & $\phantom{00.00}$ & $\phantom{00.00}$  \\
\cmidrule{1-16}
Front end  &   \multicolumn{3}{c|}{\texttt{HuBERT-XL}, fixed} &     \multicolumn{3}{c|}{\texttt{W2V-XLSR}, fixed} &    \multicolumn{3}{c|}{\texttt{W2V-Large2}, fixed}  &    \multicolumn{3}{c|}{\texttt{W2V-Large1}, fixed} &    \multicolumn{3}{c|}{\texttt{W2V-Small}, fixed} &    \multicolumn{3}{c}{}  \\
\cmidrule{1-16}
Back end &  \multicolumn{15}{c|}{\BNI} & \multicolumn{3}{c}{} \\
\cmidrule{1-16}
   &   I   &  II   &  III  &   I   &  II   &  III  &   I   &  II   &  III  &   I   &  II   &  III &   I   &  II   &  III & \multicolumn{3}{c}{} \\
\cmidrule{1-16}
   2019 LA    & \cellcolor[rgb]{0.98, 0.98, 0.98} 3.55 & \cellcolor[rgb]{0.98, 0.98, 0.98} 4.04 & \cellcolor[rgb]{0.97, 0.97, 0.97} 5.93 & \cellcolor[rgb]{0.99, 0.99, 0.99} 1.47 & \cellcolor[rgb]{0.98, 0.98, 0.98} 3.45 & \cellcolor[rgb]{0.98, 0.98, 0.98} 3.77 & \cellcolor[rgb]{1.00, 1.00, 1.00} 0.86 & \cellcolor[rgb]{1.00, 1.00, 1.00} 0.99 & \cellcolor[rgb]{0.99, 0.99, 0.99} 2.08 & \cellcolor[rgb]{0.98, 0.98, 0.98} 4.47 & \cellcolor[rgb]{0.97, 0.97, 0.97} 5.67 & \cellcolor[rgb]{0.97, 0.97, 0.97} 6.36 & \cellcolor[rgb]{0.99, 0.99, 0.99} 2.61 & \cellcolor[rgb]{0.98, 0.98, 0.98} 3.48 & \cellcolor[rgb]{0.98, 0.98, 0.98} 4.01\\ 
   2015 LA    & \cellcolor[rgb]{0.98, 0.98, 0.98} 3.27 & \cellcolor[rgb]{0.98, 0.98, 0.98} 3.25 & \cellcolor[rgb]{0.98, 0.98, 0.98} 3.69 & \cellcolor[rgb]{0.98, 0.98, 0.98} 3.97 & \cellcolor[rgb]{0.96, 0.96, 0.96} 6.78 & \cellcolor[rgb]{0.96, 0.96, 0.96} 8.18 & \cellcolor[rgb]{0.99, 0.99, 0.99} 1.39 & \cellcolor[rgb]{0.99, 0.99, 0.99} 1.39 & \cellcolor[rgb]{0.99, 0.99, 0.99} 1.99 & \cellcolor[rgb]{0.86, 0.86, 0.86} 19.66 & \cellcolor[rgb]{0.84, 0.84, 0.84} 22.33 & \cellcolor[rgb]{0.83, 0.83, 0.83} 23.65 & \cellcolor[rgb]{0.94, 0.94, 0.94} 10.40 & \cellcolor[rgb]{0.96, 0.96, 0.96} 7.58 & \cellcolor[rgb]{0.95, 0.95, 0.95} 9.28\\ 
2021 LA prog. & \cellcolor[rgb]{0.96, 0.96, 0.96} 7.63 & \cellcolor[rgb]{0.96, 0.96, 0.96} 6.61 & \cellcolor[rgb]{0.95, 0.95, 0.95} 9.55 & \cellcolor[rgb]{0.95, 0.95, 0.95} 9.85 & \cellcolor[rgb]{0.88, 0.88, 0.88} 17.29 & \cellcolor[rgb]{0.86, 0.86, 0.86} 20.17 & \cellcolor[rgb]{0.94, 0.94, 0.94} 11.40 & \cellcolor[rgb]{0.94, 0.94, 0.94} 10.50 & \cellcolor[rgb]{0.94, 0.94, 0.94} 10.92 & \cellcolor[rgb]{0.88, 0.88, 0.88} 18.25 & \cellcolor[rgb]{0.85, 0.85, 0.85} 21.00 & \cellcolor[rgb]{0.84, 0.84, 0.84} 22.32 & \cellcolor[rgb]{0.86, 0.86, 0.86} 20.28 & \cellcolor[rgb]{0.87, 0.87, 0.87} 18.91 & \cellcolor[rgb]{0.88, 0.88, 0.88} 18.57\\ 
2021 LA eval. & \cellcolor[rgb]{0.95, 0.95, 0.95} 9.55 & \cellcolor[rgb]{0.96, 0.96, 0.96} 7.03 & \cellcolor[rgb]{0.94, 0.94, 0.94} 10.54 & \cellcolor[rgb]{0.94, 0.94, 0.94} 10.97 & \cellcolor[rgb]{0.87, 0.87, 0.87} 18.91 & \cellcolor[rgb]{0.86, 0.86, 0.86} 20.71 & \cellcolor[rgb]{0.92, 0.92, 0.92} 13.19 & \cellcolor[rgb]{0.93, 0.93, 0.93} 12.57 & \cellcolor[rgb]{0.92, 0.92, 0.92} 12.94 & \cellcolor[rgb]{0.92, 0.92, 0.92} 13.86 & \cellcolor[rgb]{0.89, 0.89, 0.89} 16.77 & \cellcolor[rgb]{0.87, 0.87, 0.87} 19.38 & \cellcolor[rgb]{0.90, 0.90, 0.90} 16.11 & \cellcolor[rgb]{0.91, 0.91, 0.91} 14.79 & \cellcolor[rgb]{0.90, 0.90, 0.90} 15.56\\ 
2021 DF prog. & \cellcolor[rgb]{0.98, 0.98, 0.98} 4.16 & \cellcolor[rgb]{0.98, 0.98, 0.98} 4.32 & \cellcolor[rgb]{0.97, 0.97, 0.97} 5.11 & \cellcolor[rgb]{0.99, 0.99, 0.99} 2.67 & \cellcolor[rgb]{0.97, 0.97, 0.97} 5.09 & \cellcolor[rgb]{0.96, 0.96, 0.96} 7.02 & \cellcolor[rgb]{0.99, 0.99, 0.99} 1.86 & \cellcolor[rgb]{0.99, 0.99, 0.99} 2.12 & \cellcolor[rgb]{0.98, 0.98, 0.98} 3.36 & \cellcolor[rgb]{0.96, 0.96, 0.96} 8.22 & \cellcolor[rgb]{0.94, 0.94, 0.94} 10.32 & \cellcolor[rgb]{0.92, 0.92, 0.92} 12.92 & \cellcolor[rgb]{0.97, 0.97, 0.97} 5.34 & \cellcolor[rgb]{0.96, 0.96, 0.96} 7.80 & \cellcolor[rgb]{0.96, 0.96, 0.96} 7.87\\ 
2021 DF eval. & \cellcolor[rgb]{0.92, 0.92, 0.92} 13.07 & \cellcolor[rgb]{0.92, 0.92, 0.92} 12.87 & \cellcolor[rgb]{0.93, 0.93, 0.93} 12.39 & \cellcolor[rgb]{0.96, 0.96, 0.96} 7.14 & \cellcolor[rgb]{0.95, 0.95, 0.95} 9.94 & \cellcolor[rgb]{0.94, 0.94, 0.94} 11.35 & \cellcolor[rgb]{0.96, 0.96, 0.96} 7.44 & \cellcolor[rgb]{0.96, 0.96, 0.96} 7.77 & \cellcolor[rgb]{0.95, 0.95, 0.95} 9.26 & \cellcolor[rgb]{0.87, 0.87, 0.87} 19.26 & \cellcolor[rgb]{0.87, 0.87, 0.87} 18.68 & \cellcolor[rgb]{0.86, 0.86, 0.86} 20.75 & \cellcolor[rgb]{0.88, 0.88, 0.88} 17.74 & \cellcolor[rgb]{0.89, 0.89, 0.89} 17.00 & \cellcolor[rgb]{0.87, 0.87, 0.87} 18.97\\ 
\cmidrule{1-16}
\end{tabular}
}
\end{center}
\vspace{-5mm}
\label{tab:result}
\end{table*}

\section{Experiment}
\label{sec:exp}

\subsection{Databases and protocols}
\label{sec:data}
The training set of the ASVspoof 2019 LA database \cite{Todisco2019} was to train the CMs, and the development set was used for early stopping. 
Each CM was then evaluated on multiple test sets, including the test sets from ASVspoof 2019 LA,  2015 \cite{wu2015asvspoof}, 2021 LA, and 2021 deepfake (DF) scenarios \cite{yamagishi21_asvspoof}. 
Using the LA 2019 test set measures the CM's performance in a benign condition,
while the 2021 LA and DF test sets simulate more adverse scenarios where most of the spoofed and bona fide trials were compressed using codecs \cite{yamagishi21_asvspoof}. The DF evaluation track is more challenging because many trials are from a mismatched domain or produced by more diverse spoofing attackers. 

Evaluation on the 2021 LA and DF test sets measures the CM's generalizability to unseen attacks and unknown domains. 
The ASVspoof challenge 2015 test set is theoretically easy if a CM is trained on the advanced LA 2019 train set, but an empirical study suggests not so \cite{das2020assessing}. 
Therefore, this test set is also used in this study.

\subsection{Model configurations and training recipes}
\label{sec:recipe}
We followed our previous study to configure the CMs \cite{wang2021comparative}. 
The baseline used LFCC extracted with a frame length of 20 ms, a frame shift of 10 ms, and a 512-point FFT. The LFCC vector per frame had 60 dimensions, including static, delta, and delta-delta components. The baseline back end is plotted in Figure~\ref{fig:idea}. The training recipe was borrowed from our previous study: the Adam optimizer with $\beta_1=0.9, \beta_2=0.999, \epsilon=10^{-8}$ \cite{kingma2014adam},  a mini-batch size of 64, and a learning rate initialized to $3\times10^{-4}$ and halved every ten epochs.  

For other CMs, the front end was initialized using one of the pre-trained models in Table~\ref{tab:ssl}.
If the fine-tuning strategy was used, the front end was updated jointly with the rest of the CM on the 2019 LA training set. 
The mini-batch size was set to 8, and the learning rate was reduced\footnote{Using the same learning rate as the baseline caused overfitting} to $1\times{10}^{-6}$. Masking was not applied on the hidden feature $\bs{z}_{1:N}$ inside the Wav2vec 2.0 during fine-tuning.
If the front end was not fine-tuned, the CM was trained in the same manner as the baseline. 
The FC layer after the self-supervised front end used 128 output dimensions. 

Because of the increased GPU memory consumption when fine-tuning a self-supervised model,  the input trials during training were sliced into segments with a maximum duration of 4 s. 
The same strategy was applied to all the experimental CMs\footnote{Accordingly, the baseline was re-trained, but the results are similar to those in our previous study \cite{wang2021comparative}. The baseline is also different from the ASVspoof 2021 baseline as the LFCC configuration is different.}. During inference, however, the input trial was processed as a whole.  Voice activity detection and feature normalization were not applied. 

We trained each CM for three rounds, where each round used a different random seed to initialize the network weights (except for the pre-trained super-supervised front end). The weight initialization strategy was the default one in the Pytorch toolkit \cite{NEURIPS2019_9015}. We evaluated the three trained `versions'  separately on the test sets. 

\begin{table*}[t]
\caption{Decomposed EERs (\%) on the test sets. EERs from the three training-evaluation rounds are averaged for each model. Results on 2021 LA and DF are decomposed over codecs.}
\vspace{-5mm}
\begin{center}
\resizebox{\textwidth}{!}
{
\setlength{\tabcolsep}{2pt}
\begin{tabular}{cccccccccccccccccccccccccc}
\toprule
\multicolumn{3}{c}{CM config}  & &  \multicolumn{13}{c}{ASVspoof 2019 LA test set (2019 LA)}  & & \multicolumn{7}{c}{ASVspoof 2021 LA test set (2021 LA)}  \\  
\cmidrule(lr){1-3}\cmidrule(lr){5-17}\cmidrule(l){19-25}
\multicolumn{2}{c}{\multirow{2}{*}{Front end}} & \multirow{2}{*}{{Back end}}& & \multicolumn{2}{c}{Known attack} & \multicolumn{4}{c}{Partially known attack} & \multicolumn{7}{c}{Unknown attack} & & \multirow{2}{*}{\rotatebox[origin=c]{0}{LA-C1}}   &  \multirow{2}{*}{\rotatebox[origin=c]{0}{LA-C2}}   &  \multirow{2}{*}{\rotatebox[origin=c]{0}{LA-C3}}   &  \multirow{2}{*}{\rotatebox[origin=c]{0}{LA-C4}}   &  \multirow{2}{*}{\rotatebox[origin=c]{0}{LA-C5}}   &  \multirow{2}{*}{\rotatebox[origin=c]{0}{LA-C6}}   &  \multirow{2}{*}{\rotatebox[origin=c]{0}{LA-C7}} \\
\cmidrule(lr){5-6}\cmidrule(lr){7-10}\cmidrule(lr){11-17} 
& &  &  &  A16  &  A19  &  A07  &  A08  &  A09  &  A17  &  A10  &  A11  &  A12  &  A13  &  A14  &  A15  &  A18 & & \\  
\midrule
LFCC & -- & \BNIshort\                                                       &  & \cellcolor[rgb]{0.98, 0.98, 0.98} 0.39 & \cellcolor[rgb]{0.94, 0.94, 0.94} 1.93 & \cellcolor[rgb]{0.97, 0.97, 0.97} 0.61 & \cellcolor[rgb]{0.99, 0.99, 0.99} 0.07 & \cellcolor[rgb]{1.00, 1.00, 1.00} 0.02 & \cellcolor[rgb]{0.78, 0.78, 0.78} 14.49 & \cellcolor[rgb]{0.95, 0.95, 0.95} 1.37 & \cellcolor[rgb]{0.99, 0.99, 0.99} 0.17 & \cellcolor[rgb]{0.95, 0.95, 0.95} 1.50 & \cellcolor[rgb]{0.96, 0.96, 0.96} 1.21 & \cellcolor[rgb]{0.98, 0.98, 0.98} 0.23 & \cellcolor[rgb]{0.97, 0.97, 0.97} 0.50 & \cellcolor[rgb]{0.93, 0.93, 0.93} 2.52 &  & \cellcolor[rgb]{0.93, 0.93, 0.93} 2.80 & \cellcolor[rgb]{0.78, 0.78, 0.78} 14.19 & \cellcolor[rgb]{0.77, 0.77, 0.77} 15.80 & \cellcolor[rgb]{0.88, 0.88, 0.88} 5.65 & \cellcolor[rgb]{0.78, 0.78, 0.78} 14.10 & \cellcolor[rgb]{0.79, 0.79, 0.79} 13.65 & \cellcolor[rgb]{0.59, 0.59, 0.59} 50.30\\  
\midrule
\multirow{5}{*}{\texttt{W2V-XLSR}} & \multirow{3}{*}{{Fixed}} & \BNIshort\  &  & \cellcolor[rgb]{0.95, 0.95, 0.95} 1.54 & \cellcolor[rgb]{0.92, 0.92, 0.92} 3.47 & \cellcolor[rgb]{0.95, 0.95, 0.95} 1.69 & \cellcolor[rgb]{0.94, 0.94, 0.94} 2.12 & \cellcolor[rgb]{0.96, 0.96, 0.96} 0.98 & \cellcolor[rgb]{0.96, 0.96, 0.96} 0.93 & \cellcolor[rgb]{0.84, 0.84, 0.84} 8.67 & \cellcolor[rgb]{0.93, 0.93, 0.93} 2.64 & \cellcolor[rgb]{0.94, 0.94, 0.94} 2.10 & \cellcolor[rgb]{0.97, 0.97, 0.97} 0.64 & \cellcolor[rgb]{0.95, 0.95, 0.95} 1.53 & \cellcolor[rgb]{0.92, 0.92, 0.92} 2.94 & \cellcolor[rgb]{0.95, 0.95, 0.95} 1.48 &  & \cellcolor[rgb]{0.92, 0.92, 0.92} 2.94 & \cellcolor[rgb]{0.81, 0.81, 0.81} 12.03 & \cellcolor[rgb]{0.70, 0.70, 0.70} 24.26 & \cellcolor[rgb]{0.91, 0.91, 0.91} 3.69 & \cellcolor[rgb]{0.81, 0.81, 0.81} 11.71 & \cellcolor[rgb]{0.74, 0.74, 0.74} 19.40 & \cellcolor[rgb]{0.88, 0.88, 0.88} 5.55\\ 
& & \BNIIshort\ &                                                              & \cellcolor[rgb]{0.91, 0.91, 0.91} 3.58 & \cellcolor[rgb]{0.80, 0.80, 0.80} 12.58 & \cellcolor[rgb]{0.92, 0.92, 0.92} 3.04 & \cellcolor[rgb]{0.92, 0.92, 0.92} 3.21 & \cellcolor[rgb]{0.91, 0.91, 0.91} 3.72 & \cellcolor[rgb]{0.91, 0.91, 0.91} 3.78 & \cellcolor[rgb]{0.81, 0.81, 0.81} 11.13 & \cellcolor[rgb]{0.80, 0.80, 0.80} 12.28 & \cellcolor[rgb]{0.92, 0.92, 0.92} 3.33 & \cellcolor[rgb]{0.94, 0.94, 0.94} 2.14 & \cellcolor[rgb]{0.92, 0.92, 0.92} 3.47 & \cellcolor[rgb]{0.86, 0.86, 0.86} 6.81 & \cellcolor[rgb]{0.93, 0.93, 0.93} 2.79 &  & \cellcolor[rgb]{0.86, 0.86, 0.86} 7.06 & \cellcolor[rgb]{0.79, 0.79, 0.79} 13.65 & \cellcolor[rgb]{0.68, 0.68, 0.68} 25.90 & \cellcolor[rgb]{0.86, 0.86, 0.86} 7.46 & \cellcolor[rgb]{0.80, 0.80, 0.80} 12.56 & \cellcolor[rgb]{0.66, 0.66, 0.66} 29.34 & \cellcolor[rgb]{0.83, 0.83, 0.83} 9.58\\ 
& & \BNIIIshort\ &                                                             & \cellcolor[rgb]{0.78, 0.78, 0.78} 14.03 & \cellcolor[rgb]{0.61, 0.61, 0.61} 39.74 & \cellcolor[rgb]{0.84, 0.84, 0.84} 8.89 & \cellcolor[rgb]{0.79, 0.79, 0.79} 13.34 & \cellcolor[rgb]{0.85, 0.85, 0.85} 8.36 & \cellcolor[rgb]{0.77, 0.77, 0.77} 15.59 & \cellcolor[rgb]{0.79, 0.79, 0.79} 13.25 & \cellcolor[rgb]{0.84, 0.84, 0.84} 8.65 & \cellcolor[rgb]{0.86, 0.86, 0.86} 7.57 & \cellcolor[rgb]{0.94, 0.94, 0.94} 1.94 & \cellcolor[rgb]{0.70, 0.70, 0.70} 22.97 & \cellcolor[rgb]{0.65, 0.65, 0.65} 30.52 & \cellcolor[rgb]{0.77, 0.77, 0.77} 15.64 &  & \cellcolor[rgb]{0.74, 0.74, 0.74} 18.98 & \cellcolor[rgb]{0.72, 0.72, 0.72} 21.94 & \cellcolor[rgb]{0.71, 0.71, 0.71} 22.72 & \cellcolor[rgb]{0.74, 0.74, 0.74} 19.05 & \cellcolor[rgb]{0.72, 0.72, 0.72} 20.85 & \cellcolor[rgb]{0.70, 0.70, 0.70} 24.17 & \cellcolor[rgb]{0.75, 0.75, 0.75} 17.39\\  
\cmidrule{2-25} 
& \multirow{3}{*}{\shortstack{Fine- \\ tuned}} & \BNIshort\  &                  & \cellcolor[rgb]{0.99, 0.99, 0.99} 0.07 & \cellcolor[rgb]{0.99, 0.99, 0.99} 0.20 & \cellcolor[rgb]{0.99, 0.99, 0.99} 0.16 & \cellcolor[rgb]{0.99, 0.99, 0.99} 0.18 & \cellcolor[rgb]{0.99, 0.99, 0.99} 0.19 & \cellcolor[rgb]{0.99, 0.99, 0.99} 0.12 & \cellcolor[rgb]{0.87, 0.87, 0.87} 6.28 & \cellcolor[rgb]{0.80, 0.80, 0.80} 12.86 & \cellcolor[rgb]{0.99, 0.99, 0.99} 0.17 & \cellcolor[rgb]{1.00, 1.00, 1.00} 0.04 & \cellcolor[rgb]{0.97, 0.97, 0.97} 0.49 & \cellcolor[rgb]{0.95, 0.95, 0.95} 1.58 & \cellcolor[rgb]{0.98, 0.98, 0.98} 0.26 &   & \cellcolor[rgb]{0.92, 0.92, 0.92} 3.16 & \cellcolor[rgb]{0.90, 0.90, 0.90} 4.55 & \cellcolor[rgb]{0.83, 0.83, 0.83} 9.93 & \cellcolor[rgb]{0.91, 0.91, 0.91} 3.84 & \cellcolor[rgb]{0.90, 0.90, 0.90} 4.38 & \cellcolor[rgb]{0.85, 0.85, 0.85} 8.50 & \cellcolor[rgb]{0.88, 0.88, 0.88} 6.07\\ 
& &  \BNIIshort\ &                                                             & \cellcolor[rgb]{0.99, 0.99, 0.99} 0.11 & \cellcolor[rgb]{0.99, 0.99, 0.99} 0.17 & \cellcolor[rgb]{0.99, 0.99, 0.99} 0.12 & \cellcolor[rgb]{0.99, 0.99, 0.99} 0.14 & \cellcolor[rgb]{0.99, 0.99, 0.99} 0.07 & \cellcolor[rgb]{0.99, 0.99, 0.99} 0.05 & \cellcolor[rgb]{0.91, 0.91, 0.91} 3.58 & \cellcolor[rgb]{0.92, 0.92, 0.92} 3.06 & \cellcolor[rgb]{0.99, 0.99, 0.99} 0.12 & \cellcolor[rgb]{1.00, 1.00, 1.00} 0.02 & \cellcolor[rgb]{0.99, 0.99, 0.99} 0.18 & \cellcolor[rgb]{0.96, 0.96, 0.96} 0.97 & \cellcolor[rgb]{0.98, 0.98, 0.98} 0.23 &   & \cellcolor[rgb]{0.94, 0.94, 0.94} 2.14 & \cellcolor[rgb]{0.90, 0.90, 0.90} 4.41 & \cellcolor[rgb]{0.83, 0.83, 0.83} 9.78 & \cellcolor[rgb]{0.92, 0.92, 0.92} 3.29 & \cellcolor[rgb]{0.90, 0.90, 0.90} 4.05 & \cellcolor[rgb]{0.87, 0.87, 0.87} 6.21 & \cellcolor[rgb]{0.89, 0.89, 0.89} 5.08\\
& &  \BNIIIshort\ &                                                            & \cellcolor[rgb]{0.99, 0.99, 0.99} 0.10 & \cellcolor[rgb]{0.99, 0.99, 0.99} 0.14 & \cellcolor[rgb]{0.99, 0.99, 0.99} 0.16 & \cellcolor[rgb]{0.99, 0.99, 0.99} 0.14 & \cellcolor[rgb]{0.99, 0.99, 0.99} 0.17 & \cellcolor[rgb]{0.99, 0.99, 0.99} 0.08 & \cellcolor[rgb]{0.89, 0.89, 0.89} 5.15 & \cellcolor[rgb]{0.88, 0.88, 0.88} 5.59 & \cellcolor[rgb]{0.99, 0.99, 0.99} 0.17 & \cellcolor[rgb]{0.99, 0.99, 0.99} 0.06 & \cellcolor[rgb]{0.98, 0.98, 0.98} 0.26 & \cellcolor[rgb]{0.96, 0.96, 0.96} 1.09 & \cellcolor[rgb]{0.99, 0.99, 0.99} 0.18 &   & \cellcolor[rgb]{0.92, 0.92, 0.92} 3.25 & \cellcolor[rgb]{0.89, 0.89, 0.89} 5.04 & \cellcolor[rgb]{0.82, 0.82, 0.82} 10.79 & \cellcolor[rgb]{0.90, 0.90, 0.90} 4.16 & \cellcolor[rgb]{0.89, 0.89, 0.89} 4.77 & \cellcolor[rgb]{0.84, 0.84, 0.84} 8.81 & \cellcolor[rgb]{0.88, 0.88, 0.88} 5.45\\  
\midrule
\texttt{HuBERT-XL} & \multirow{4}{*}{{Fixed}}  & \multirow{4}{*}{{\BNIshort}}& & \cellcolor[rgb]{0.95, 0.95, 0.95} 1.47 & \cellcolor[rgb]{0.82, 0.82, 0.82} 10.66 & \cellcolor[rgb]{0.95, 0.95, 0.95} 1.68 & \cellcolor[rgb]{0.92, 0.92, 0.92} 3.16 & \cellcolor[rgb]{0.97, 0.97, 0.97} 0.63 & \cellcolor[rgb]{0.96, 0.96, 0.96} 1.13 & \cellcolor[rgb]{0.84, 0.84, 0.84} 9.10 & \cellcolor[rgb]{0.93, 0.93, 0.93} 2.87 & \cellcolor[rgb]{0.97, 0.97, 0.97} 0.59 & \cellcolor[rgb]{0.99, 0.99, 0.99} 0.09 & \cellcolor[rgb]{0.94, 0.94, 0.94} 2.37 & \cellcolor[rgb]{0.94, 0.94, 0.94} 2.09 & \cellcolor[rgb]{0.85, 0.85, 0.85} 8.22 &   & \cellcolor[rgb]{0.90, 0.90, 0.90} 4.68 & \cellcolor[rgb]{0.88, 0.88, 0.88} 5.35 & \cellcolor[rgb]{0.87, 0.87, 0.87} 6.54 & \cellcolor[rgb]{0.90, 0.90, 0.90} 4.56 & \cellcolor[rgb]{0.88, 0.88, 0.88} 5.40 & \cellcolor[rgb]{0.86, 0.86, 0.86} 7.33 & \cellcolor[rgb]{0.90, 0.90, 0.90} 4.68\\ 
\texttt{W2V-Large2} & &  &                                                     & \cellcolor[rgb]{0.97, 0.97, 0.97} 0.65 & \cellcolor[rgb]{0.93, 0.93, 0.93} 2.79 & \cellcolor[rgb]{0.96, 0.96, 0.96} 0.88 & \cellcolor[rgb]{0.96, 0.96, 0.96} 1.27 & \cellcolor[rgb]{0.98, 0.98, 0.98} 0.23 & \cellcolor[rgb]{0.98, 0.98, 0.98} 0.33 & \cellcolor[rgb]{0.93, 0.93, 0.93} 2.80 & \cellcolor[rgb]{0.97, 0.97, 0.97} 0.59 & \cellcolor[rgb]{0.97, 0.97, 0.97} 0.61 & \cellcolor[rgb]{0.99, 0.99, 0.99} 0.08 & \cellcolor[rgb]{0.97, 0.97, 0.97} 0.64 & \cellcolor[rgb]{0.96, 0.96, 0.96} 0.84 & \cellcolor[rgb]{0.98, 0.98, 0.98} 0.36 &   & \cellcolor[rgb]{0.96, 0.96, 0.96} 1.30 & \cellcolor[rgb]{0.91, 0.91, 0.91} 3.86 & \cellcolor[rgb]{0.80, 0.80, 0.80} 12.61 & \cellcolor[rgb]{0.94, 0.94, 0.94} 2.01 & \cellcolor[rgb]{0.91, 0.91, 0.91} 4.03 & \cellcolor[rgb]{0.79, 0.79, 0.79} 13.24 & \cellcolor[rgb]{0.92, 0.92, 0.92} 2.93\\ 
\texttt{W2V-Large1} & &  &                                                     & \cellcolor[rgb]{0.90, 0.90, 0.90} 4.29 & \cellcolor[rgb]{0.84, 0.84, 0.84} 8.65 & \cellcolor[rgb]{0.89, 0.89, 0.89} 5.18 & \cellcolor[rgb]{0.90, 0.90, 0.90} 4.10 & \cellcolor[rgb]{0.93, 0.93, 0.93} 2.55 & \cellcolor[rgb]{0.86, 0.86, 0.86} 7.54 & \cellcolor[rgb]{0.84, 0.84, 0.84} 8.89 & \cellcolor[rgb]{0.91, 0.91, 0.91} 3.78 & \cellcolor[rgb]{0.93, 0.93, 0.93} 2.90 & \cellcolor[rgb]{0.93, 0.93, 0.93} 2.65 & \cellcolor[rgb]{0.89, 0.89, 0.89} 4.87 & \cellcolor[rgb]{0.88, 0.88, 0.88} 5.76 & \cellcolor[rgb]{0.87, 0.87, 0.87} 6.65 &  & \cellcolor[rgb]{0.88, 0.88, 0.88} 5.97 & \cellcolor[rgb]{0.84, 0.84, 0.84} 8.62 & \cellcolor[rgb]{0.61, 0.61, 0.61} 40.56 & \cellcolor[rgb]{0.86, 0.86, 0.86} 7.35 & \cellcolor[rgb]{0.85, 0.85, 0.85} 8.15 & \cellcolor[rgb]{0.73, 0.73, 0.73} 19.63 & \cellcolor[rgb]{0.86, 0.86, 0.86} 7.24\\ 
\texttt{W2V-Small} & &  &                                                      & \cellcolor[rgb]{0.93, 0.93, 0.93} 2.52 & \cellcolor[rgb]{0.89, 0.89, 0.89} 5.21 & \cellcolor[rgb]{0.91, 0.91, 0.91} 3.53 & \cellcolor[rgb]{0.95, 0.95, 0.95} 1.52 & \cellcolor[rgb]{0.97, 0.97, 0.97} 0.59 & \cellcolor[rgb]{0.88, 0.88, 0.88} 5.42 & \cellcolor[rgb]{0.88, 0.88, 0.88} 5.77 & \cellcolor[rgb]{0.96, 0.96, 0.96} 1.30 & \cellcolor[rgb]{0.96, 0.96, 0.96} 1.16 & \cellcolor[rgb]{0.96, 0.96, 0.96} 0.92 & \cellcolor[rgb]{0.95, 0.95, 0.95} 1.80 & \cellcolor[rgb]{0.93, 0.93, 0.93} 2.72 & \cellcolor[rgb]{0.93, 0.93, 0.93} 2.63 &   & \cellcolor[rgb]{0.90, 0.90, 0.90} 4.05 & \cellcolor[rgb]{0.87, 0.87, 0.87} 6.64 & \cellcolor[rgb]{0.72, 0.72, 0.72} 21.29 & \cellcolor[rgb]{0.88, 0.88, 0.88} 5.86 & \cellcolor[rgb]{0.87, 0.87, 0.87} 6.38 & \cellcolor[rgb]{0.80, 0.80, 0.80} 12.89 & \cellcolor[rgb]{0.90, 0.90, 0.90} 4.24\\
\midrule
\multicolumn{3}{c}{CM config}  & &  \multicolumn{10}{c}{ASVspoof 2015 test set (2015 LA)}  & & & \multicolumn{9}{c}{ASVspoof 2021 DF test set (2021 DF)}  \\  
\cmidrule(lr){1-3}\cmidrule(lr){5-14}\cmidrule(lr){17-25}
\multicolumn{2}{c}{{Front end}} & {{Back end}}  & &   S1   &  S2   &  S3   &  S4   &  S5   &  S6   &  S7   &  S8   &  S9   &  S10  &  & & \rotatebox[origin=c]{0}{DF-C1} &  \rotatebox[origin=c]{0}{DF-C2} & \rotatebox[origin=c]{0}{DF-C3} & \rotatebox[origin=c]{0}{DF-C4} & \rotatebox[origin=c]{0}{DF-C5} & \rotatebox[origin=c]{0}{DF-C6} & \rotatebox[origin=c]{0}{DF-C7} & \rotatebox[origin=c]{0}{DF-C8} & \rotatebox[origin=c]{0}{DF-C9} \\  
\midrule
LFCC & -- & \BNIshort\                                                       &   & \cellcolor[rgb]{0.59, 0.59, 0.59} 49.68 & \cellcolor[rgb]{0.59, 0.59, 0.59} 46.26 & \cellcolor[rgb]{0.82, 0.82, 0.82} 10.87 & \cellcolor[rgb]{0.84, 0.84, 0.84} 9.17 & \cellcolor[rgb]{0.66, 0.66, 0.66} 28.83 & \cellcolor[rgb]{0.65, 0.65, 0.65} 30.06 & \cellcolor[rgb]{0.83, 0.83, 0.83} 9.44 & \cellcolor[rgb]{0.66, 0.66, 0.66} 28.32 & \cellcolor[rgb]{0.82, 0.82, 0.82} 10.58 & \cellcolor[rgb]{0.61, 0.61, 0.61} 38.84 &  &  & \cellcolor[rgb]{0.85, 0.85, 0.85} 18.94 & \cellcolor[rgb]{0.67, 0.67, 0.67} 38.56 & \cellcolor[rgb]{0.61, 0.61, 0.61} 46.38 & \cellcolor[rgb]{0.83, 0.83, 0.83} 20.35 & \cellcolor[rgb]{0.84, 0.84, 0.84} 19.32 & \cellcolor[rgb]{0.85, 0.85, 0.85} 18.46 & \cellcolor[rgb]{0.89, 0.89, 0.89} 13.25 & \cellcolor[rgb]{0.74, 0.74, 0.74} 31.61 & \cellcolor[rgb]{0.86, 0.86, 0.86} 17.15\\ 
\midrule
\multirow{5}{*}{\texttt{W2V-XLSR}} & \multirow{3}{*}{{Fixed}} & \BNIshort\  &   & \cellcolor[rgb]{0.98, 0.98, 0.98} 0.35 & \cellcolor[rgb]{0.77, 0.77, 0.77} 15.55 & \cellcolor[rgb]{0.86, 0.86, 0.86} 7.71 & \cellcolor[rgb]{0.86, 0.86, 0.86} 7.23 & \cellcolor[rgb]{0.92, 0.92, 0.92} 3.13 & \cellcolor[rgb]{0.90, 0.90, 0.90} 4.48 & \cellcolor[rgb]{0.91, 0.91, 0.91} 3.76 & \cellcolor[rgb]{0.94, 0.94, 0.94} 2.47 & \cellcolor[rgb]{0.94, 0.94, 0.94} 1.90 & \cellcolor[rgb]{0.88, 0.88, 0.88} 5.65 &  &  & \cellcolor[rgb]{0.93, 0.93, 0.93} 8.98 & \cellcolor[rgb]{0.90, 0.90, 0.90} 12.31 & \cellcolor[rgb]{0.92, 0.92, 0.92} 10.72 & \cellcolor[rgb]{0.93, 0.93, 0.93} 9.04 & \cellcolor[rgb]{0.93, 0.93, 0.93} 8.95 & \cellcolor[rgb]{0.94, 0.94, 0.94} 7.42 & \cellcolor[rgb]{0.95, 0.95, 0.95} 6.38 & \cellcolor[rgb]{0.92, 0.92, 0.92} 10.05 & \cellcolor[rgb]{0.94, 0.94, 0.94} 7.27\\ 
& & \BNIIshort\ &                                                               & \cellcolor[rgb]{0.91, 0.91, 0.91} 3.92 & \cellcolor[rgb]{0.77, 0.77, 0.77} 15.10 & \cellcolor[rgb]{0.79, 0.79, 0.79} 13.65 & \cellcolor[rgb]{0.78, 0.78, 0.78} 13.80 & \cellcolor[rgb]{0.87, 0.87, 0.87} 6.44 & \cellcolor[rgb]{0.86, 0.86, 0.86} 7.34 & \cellcolor[rgb]{0.86, 0.86, 0.86} 7.57 & \cellcolor[rgb]{0.86, 0.86, 0.86} 7.61 & \cellcolor[rgb]{0.88, 0.88, 0.88} 5.33 & \cellcolor[rgb]{0.83, 0.83, 0.83} 10.05 &  &  & \cellcolor[rgb]{0.90, 0.90, 0.90} 12.20 & \cellcolor[rgb]{0.86, 0.86, 0.86} 17.30 & \cellcolor[rgb]{0.88, 0.88, 0.88} 14.89 & \cellcolor[rgb]{0.90, 0.90, 0.90} 13.00 & \cellcolor[rgb]{0.90, 0.90, 0.90} 12.49 & \cellcolor[rgb]{0.91, 0.91, 0.91} 11.38 & \cellcolor[rgb]{0.92, 0.92, 0.92} 10.03 & \cellcolor[rgb]{0.88, 0.88, 0.88} 14.27 & \cellcolor[rgb]{0.91, 0.91, 0.91} 11.72\\ 
& & \BNIIIshort\ &                                                              & \cellcolor[rgb]{0.92, 0.92, 0.92} 3.01 & \cellcolor[rgb]{0.63, 0.63, 0.63} 34.03 & \cellcolor[rgb]{0.76, 0.76, 0.76} 17.00 & \cellcolor[rgb]{0.76, 0.76, 0.76} 16.99 & \cellcolor[rgb]{0.77, 0.77, 0.77} 15.05 & \cellcolor[rgb]{0.71, 0.71, 0.71} 22.60 & \cellcolor[rgb]{0.71, 0.71, 0.71} 22.66 & \cellcolor[rgb]{0.87, 0.87, 0.87} 6.33 & \cellcolor[rgb]{0.80, 0.80, 0.80} 12.74 & \cellcolor[rgb]{0.82, 0.82, 0.82} 10.63 &  &  & \cellcolor[rgb]{0.84, 0.84, 0.84} 19.55 & \cellcolor[rgb]{0.83, 0.83, 0.83} 20.43 & \cellcolor[rgb]{0.84, 0.84, 0.84} 19.33 & \cellcolor[rgb]{0.83, 0.83, 0.83} 20.53 & \cellcolor[rgb]{0.84, 0.84, 0.84} 19.91 & \cellcolor[rgb]{0.85, 0.85, 0.85} 18.88 & \cellcolor[rgb]{0.85, 0.85, 0.85} 18.40 & \cellcolor[rgb]{0.84, 0.84, 0.84} 19.61 & \cellcolor[rgb]{0.84, 0.84, 0.84} 19.55\\ 
\cmidrule{2-25}
& \multirow{3}{*}{\shortstack{Fine- \\ tuned}} & \BNIshort\  &                   & \cellcolor[rgb]{0.99, 0.99, 0.99} 0.04 & \cellcolor[rgb]{0.96, 0.96, 0.96} 0.93 & \cellcolor[rgb]{0.99, 0.99, 0.99} 0.10 & \cellcolor[rgb]{0.99, 0.99, 0.99} 0.10 & \cellcolor[rgb]{0.99, 0.99, 0.99} 0.06 & \cellcolor[rgb]{0.99, 0.99, 0.99} 0.10 & \cellcolor[rgb]{0.99, 0.99, 0.99} 0.14 & \cellcolor[rgb]{0.99, 0.99, 0.99} 0.14 & \cellcolor[rgb]{0.99, 0.99, 0.99} 0.06 & \cellcolor[rgb]{0.98, 0.98, 0.98} 0.23 &  &  & \cellcolor[rgb]{0.95, 0.95, 0.95} 6.17 & \cellcolor[rgb]{0.93, 0.93, 0.93} 8.81 & \cellcolor[rgb]{0.95, 0.95, 0.95} 6.48 & \cellcolor[rgb]{0.95, 0.95, 0.95} 6.46 & \cellcolor[rgb]{0.95, 0.95, 0.95} 6.21 & \cellcolor[rgb]{0.96, 0.96, 0.96} 5.00 & \cellcolor[rgb]{0.96, 0.96, 0.96} 5.02 & \cellcolor[rgb]{0.95, 0.95, 0.95} 6.54 & \cellcolor[rgb]{0.96, 0.96, 0.96} 5.01\\ 
& &  \BNIIshort\ &                                                              & \cellcolor[rgb]{1.00, 1.00, 1.00} 0.01 & \cellcolor[rgb]{0.96, 0.96, 0.96} 0.89 & \cellcolor[rgb]{0.99, 0.99, 0.99} 0.06 & \cellcolor[rgb]{0.99, 0.99, 0.99} 0.04 & \cellcolor[rgb]{0.99, 0.99, 0.99} 0.05 & \cellcolor[rgb]{0.99, 0.99, 0.99} 0.07 & \cellcolor[rgb]{0.99, 0.99, 0.99} 0.09 & \cellcolor[rgb]{1.00, 1.00, 1.00} 0.03 & \cellcolor[rgb]{1.00, 1.00, 1.00} 0.03 & \cellcolor[rgb]{0.99, 0.99, 0.99} 0.18 &  &  & \cellcolor[rgb]{0.96, 0.96, 0.96} 5.15 & \cellcolor[rgb]{0.96, 0.96, 0.96} 5.49 & \cellcolor[rgb]{0.96, 0.96, 0.96} 5.58 & \cellcolor[rgb]{0.96, 0.96, 0.96} 5.06 & \cellcolor[rgb]{0.96, 0.96, 0.96} 5.00 & \cellcolor[rgb]{0.96, 0.96, 0.96} 4.28 & \cellcolor[rgb]{0.96, 0.96, 0.96} 4.32 & \cellcolor[rgb]{0.96, 0.96, 0.96} 4.71 & \cellcolor[rgb]{0.96, 0.96, 0.96} 4.28\\ 
& &  \BNIIIshort\ &                                                             & \cellcolor[rgb]{1.00, 1.00, 1.00} 0.03 & \cellcolor[rgb]{0.97, 0.97, 0.97} 0.58 & \cellcolor[rgb]{0.99, 0.99, 0.99} 0.04 & \cellcolor[rgb]{0.99, 0.99, 0.99} 0.04 & \cellcolor[rgb]{0.99, 0.99, 0.99} 0.07 & \cellcolor[rgb]{0.99, 0.99, 0.99} 0.06 & \cellcolor[rgb]{0.99, 0.99, 0.99} 0.09 & \cellcolor[rgb]{0.99, 0.99, 0.99} 0.06 & \cellcolor[rgb]{0.99, 0.99, 0.99} 0.09 & \cellcolor[rgb]{0.99, 0.99, 0.99} 0.18 &  &  & \cellcolor[rgb]{0.96, 0.96, 0.96} 5.36 & \cellcolor[rgb]{0.95, 0.95, 0.95} 6.75 & \cellcolor[rgb]{0.95, 0.95, 0.95} 6.35 & \cellcolor[rgb]{0.95, 0.95, 0.95} 5.70 & \cellcolor[rgb]{0.96, 0.96, 0.96} 5.36 & \cellcolor[rgb]{0.96, 0.96, 0.96} 4.60 & \cellcolor[rgb]{0.96, 0.96, 0.96} 4.32 & \cellcolor[rgb]{0.95, 0.95, 0.95} 5.76 & \cellcolor[rgb]{0.96, 0.96, 0.96} 4.78\\ 
\midrule
\texttt{HuBERT-XL} & \multirow{4}{*}{{Fixed}}  & \multirow{4}{*}{{\BNIshort}}&  & \cellcolor[rgb]{0.99, 0.99, 0.99} 0.04 & \cellcolor[rgb]{0.81, 0.81, 0.81} 11.82 & \cellcolor[rgb]{0.92, 0.92, 0.92} 2.91 & \cellcolor[rgb]{0.93, 0.93, 0.93} 2.60 & \cellcolor[rgb]{0.97, 0.97, 0.97} 0.73 & \cellcolor[rgb]{0.95, 0.95, 0.95} 1.38 & \cellcolor[rgb]{0.97, 0.97, 0.97} 0.68 & \cellcolor[rgb]{0.98, 0.98, 0.98} 0.39 & \cellcolor[rgb]{0.98, 0.98, 0.98} 0.44 & \cellcolor[rgb]{0.95, 0.95, 0.95} 1.48 &  &  & \cellcolor[rgb]{0.88, 0.88, 0.88} 15.36 & \cellcolor[rgb]{0.88, 0.88, 0.88} 15.36 & \cellcolor[rgb]{0.87, 0.87, 0.87} 15.57 & \cellcolor[rgb]{0.88, 0.88, 0.88} 15.10 & \cellcolor[rgb]{0.88, 0.88, 0.88} 15.34 & \cellcolor[rgb]{0.91, 0.91, 0.91} 10.90 & \cellcolor[rgb]{0.91, 0.91, 0.91} 10.82 & \cellcolor[rgb]{0.91, 0.91, 0.91} 10.78 & \cellcolor[rgb]{0.92, 0.92, 0.92} 10.74\\ 
\texttt{W2V-Large2} & &  &                                                      & \cellcolor[rgb]{0.99, 0.99, 0.99} 0.19 & \cellcolor[rgb]{0.88, 0.88, 0.88} 5.67 & \cellcolor[rgb]{0.97, 0.97, 0.97} 0.52 & \cellcolor[rgb]{0.97, 0.97, 0.97} 0.50 & \cellcolor[rgb]{0.98, 0.98, 0.98} 0.37 & \cellcolor[rgb]{0.97, 0.97, 0.97} 0.58 & \cellcolor[rgb]{0.97, 0.97, 0.97} 0.56 & \cellcolor[rgb]{0.98, 0.98, 0.98} 0.45 & \cellcolor[rgb]{0.98, 0.98, 0.98} 0.26 & \cellcolor[rgb]{0.96, 0.96, 0.96} 1.10 &  &  & \cellcolor[rgb]{0.94, 0.94, 0.94} 7.67 & \cellcolor[rgb]{0.93, 0.93, 0.93} 9.44 & \cellcolor[rgb]{0.94, 0.94, 0.94} 8.19 & \cellcolor[rgb]{0.94, 0.94, 0.94} 8.12 & \cellcolor[rgb]{0.94, 0.94, 0.94} 7.79 & \cellcolor[rgb]{0.95, 0.95, 0.95} 6.90 & \cellcolor[rgb]{0.96, 0.96, 0.96} 5.61 & \cellcolor[rgb]{0.94, 0.94, 0.94} 7.37 & \cellcolor[rgb]{0.95, 0.95, 0.95} 6.99\\ 
\texttt{W2V-Large1} & &  &                                                      & \cellcolor[rgb]{0.90, 0.90, 0.90} 4.11 & \cellcolor[rgb]{0.59, 0.59, 0.59} 46.61 & \cellcolor[rgb]{0.73, 0.73, 0.73} 20.18 & \cellcolor[rgb]{0.74, 0.74, 0.74} 19.12 & \cellcolor[rgb]{0.73, 0.73, 0.73} 20.17 & \cellcolor[rgb]{0.68, 0.68, 0.68} 25.34 & \cellcolor[rgb]{0.72, 0.72, 0.72} 20.59 & \cellcolor[rgb]{0.76, 0.76, 0.76} 15.95 & \cellcolor[rgb]{0.76, 0.76, 0.76} 16.82 & \cellcolor[rgb]{0.71, 0.71, 0.71} 22.82 &  &  & \cellcolor[rgb]{0.86, 0.86, 0.86} 17.84 & \cellcolor[rgb]{0.76, 0.76, 0.76} 28.89 & \cellcolor[rgb]{0.85, 0.85, 0.85} 18.62 & \cellcolor[rgb]{0.84, 0.84, 0.84} 19.14 & \cellcolor[rgb]{0.85, 0.85, 0.85} 18.07 & \cellcolor[rgb]{0.85, 0.85, 0.85} 18.08 & \cellcolor[rgb]{0.88, 0.88, 0.88} 14.77 & \cellcolor[rgb]{0.81, 0.81, 0.81} 23.16 & \cellcolor[rgb]{0.85, 0.85, 0.85} 18.52\\ 
\texttt{W2V-Small} & &  &                                                       & \cellcolor[rgb]{0.98, 0.98, 0.98} 0.38 & \cellcolor[rgb]{0.68, 0.68, 0.68} 25.98 & \cellcolor[rgb]{0.93, 0.93, 0.93} 2.74 & \cellcolor[rgb]{0.93, 0.93, 0.93} 2.72 & \cellcolor[rgb]{0.86, 0.86, 0.86} 7.05 & \cellcolor[rgb]{0.82, 0.82, 0.82} 10.30 & \cellcolor[rgb]{0.81, 0.81, 0.81} 11.47 & \cellcolor[rgb]{0.92, 0.92, 0.92} 3.33 & \cellcolor[rgb]{0.88, 0.88, 0.88} 5.58 & \cellcolor[rgb]{0.89, 0.89, 0.89} 5.21 &  &  & \cellcolor[rgb]{0.85, 0.85, 0.85} 18.11 & \cellcolor[rgb]{0.80, 0.80, 0.80} 24.64 & \cellcolor[rgb]{0.84, 0.84, 0.84} 19.50 & \cellcolor[rgb]{0.84, 0.84, 0.84} 19.31 & \cellcolor[rgb]{0.85, 0.85, 0.85} 18.87 & \cellcolor[rgb]{0.87, 0.87, 0.87} 15.56 & \cellcolor[rgb]{0.88, 0.88, 0.88} 14.11 & \cellcolor[rgb]{0.86, 0.86, 0.86} 17.78 & \cellcolor[rgb]{0.87, 0.87, 0.87} 15.57\\ 
\bottomrule
\end{tabular}
}
\end{center}
\vspace{-5mm}
\label{fig:sig_decomposed}
\end{table*}

\subsection{Results and discussion}
\label{sec:results}

Due to the limited computing resources, we did not exhaust all combinations of the self-supervised models and back ends. 
The investigated CMs and their EERs\footnote{Labels and codes to compute the EER are available in \url{https://github.com/asvspoof-challenge/2021}.} on the test sets are listed in Table~\ref{tab:result}. 
We conducted statistical analyses on the intra- and inter-model differences \cite{wang2021comparative}, and the results are plotted in the Appendix\footnote{See \url{https://arxiv.org/abs/2111.07725}.}. 

\subsubsection{Comparing CMs using a self-supervised front end}

\begin{figure}[t]
\centering
\includegraphics[trim=0 0 0 5, clip, width=0.48\textwidth]{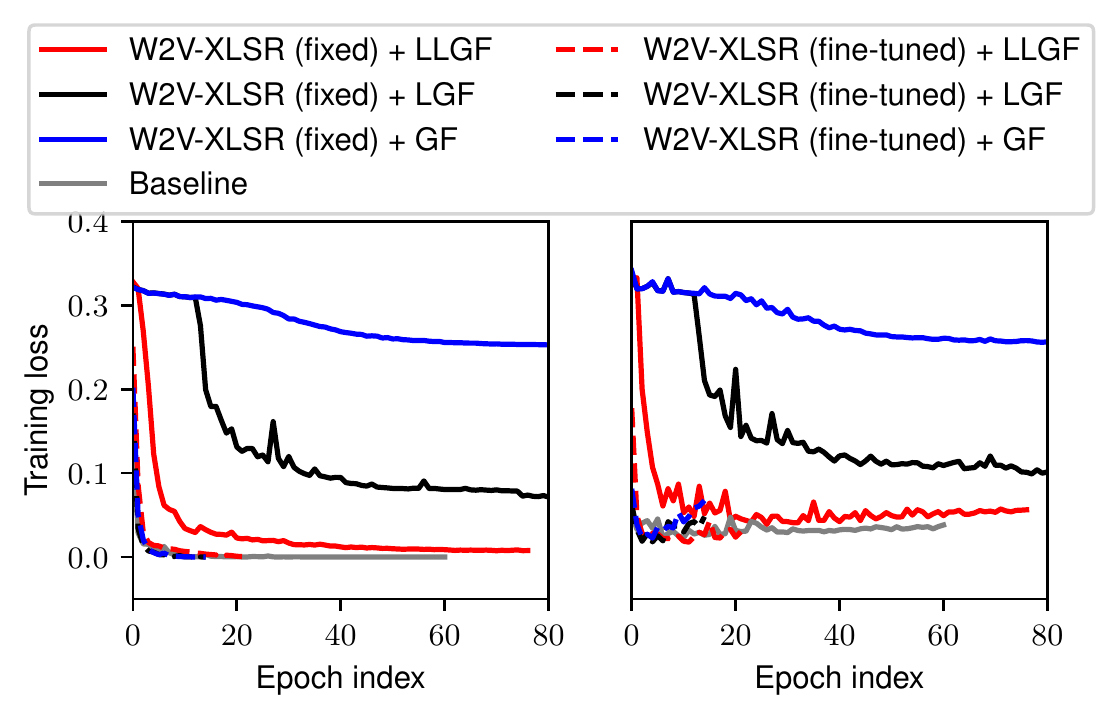}
\vspace{-8mm}
\caption{Cross-entropy loss on training (left) and development (right) sets. The best single round of each model is plotted.}
\label{fig:sig_loss}
\end{figure}

\textbf{Which back end is more suitable for the CM with a self-supervised model?} 
By comparing the three CMs using the fixed \texttt{W2V-XLSR} front end, we observe that the \BNI\ obtained lower EERs than \BNII, and \BNII\ outperformed \BNIII. Furthermore, the statistical analysis indicates that their inter-model differences are statistically significant in most cases.  Figure~\ref{fig:sig_loss} plots the learning curves on the training and development sets. By comparing the red, gray, and blue solid curves, we observe that \BNI's curve converged best. The training losses when using \BNII\ and \BNIII\  were much higher. 
The higher EERs on the test sets and training losses suggest that \BNII\ and \BNIII\ are not comparable to the deep \BNI\ when using a fixed pre-trained front end. 

However, when the front end was fine-tuned, the choice of the back end is less essential. Even the simple \BNIII\ achieved similar results on the test sets. This is discussed in the next paragraph.

\textbf{Should the self-supervised front end be fine-tuned?} 
We fine-tuned \texttt{W2V-XLSR} with the rest of the CM on the 2019 LA training set and obtained positive results on all test sets. 
No matter which back end was used, the CM with the front end fine-tuned performed similarly to or outperformed its counterpart with a fixed front end.  
The learning curves plotted in Figure~\ref{fig:sig_loss} also show that the CM converged more quickly than the case of using a fixed front end. 
Furthermore, the choice of back end has less impact on the CM performance.
Statistical analysis demonstrated that the differences between \BNI\ and \BNIII\ are not significant in most cases where the front end is fine-tuned. 

The decomposed EERs listed in Table~\ref{fig:sig_decomposed} show more notable results\footnote{
Decomposed EERs for LA 2021 and DF test sets can be obtained from the official Codalab webpages. 
LA: \url{https://competitions.codalab.org/competitions/35161}, DF: \url{https://competitions.codalab.org/competitions/35159}.}. Note that, in the 2019 LA test set, the spoofing attacks A16 and A19 can be considered as known attacks because they also produced spoofed trials for the training and development sets, even though the speakers and utterances were disjoint. Other spoofing attacks are either unknown or partially similar to the attacks in the training set. Compared with the no-fine-tuning strategy, fine-tuning the front end helped the CM improve the EERs on known and partially known attacks. Furthermore, the EERs on unknown attacks were also reduced, except for the case on A11 when using \BNI. In general, fine-tuning the self-supervised front end is worthy of trial.


\textbf{Which pre-trained self-supervised model is preferred?} 
Since fine-tuning the self-supervised front end requires more training time \footnote{Fine-tuning \texttt{HuBERT-XL} requires prohibitively more GPU memory and was not explored further.}, we fixed the pre-trained front end for this experiment.  
When combined with the back end \BNI, our results suggest that \texttt{W2V-Large2} and \texttt{W2V-XLSR} are the best two in this study. A common point on these two front ends is that both were trained using speech data from diverse corpora (see Table~\ref{tab:ssl}). The other three choices, in contrast, used data from only one corpus. 
We hypothesize that a good self-supervised front end should be trained with diverse speech data so that it can derive general and discriminative features for the anti-spoofing task across different test sets. 

\subsubsection{Comparison with the baseline}
Compared with the baseline using an LFCC-based front end, the \BNI-based CMs using fixed \texttt{W2V-XLSR} or \texttt{W2V-Large2} obtained similar or even lower EERs on all test sets. 
The CMs using the fine-tuned \texttt{W2V-XLSR} further reduced the EERs.  
More interestingly, these four CMs showed different performances from the baseline CM on individual spoofing attacks. 
For example, as Table~\ref{fig:sig_decomposed} shows,  the four CMs can detect the `most difficult attack' A17 in the 2019 LA test set with a decently low EER ($<$ 1\%). 
In contrast, A10 became the challenging attack.
Nevertheless, the four CMs performed well on most attacks.

The results on the 2021 LA and DF test sets are more notable. 
Similar to the findings in the ASVspoof 2021 challenge \cite{yamagishi21_asvspoof},  the baseline CM's performance degraded on these test sets because they contained trials processed by codec or from a different domain.
Using a pre-trained self-supervised front end alleviated the issue.
When the front end was not fine-tuned, the combination of \BNI\ with \texttt{W2V-XLSR} or \texttt{W2V-Large2} reduced the EERs on both test sets. 
When the front end was fine-tuned on the 2019 LA training set, the three CMs using different back ends all obtained much lower EERs. 
Table~\ref{fig:sig_decomposed} shows that the three CMs obtained stable results across different codecs except for LA-C3 and LA-C6 in the 2021 LA set. 
Notice how the baseline obtained a 50\% EER on LA-C7 and DF-C3. In contrast, the three CMs using a fine-tuned front end obtained EERs of less than 7\%.

On the ASVspoof 2015 test set, similar to the observations in another study \cite{das2020assessing}, the baseline CM trained on the 2019 LA training set performed poorly on this `easy' test set. 
By using a fine-tuned front end, the CMs obtained EERs of less than 1\% over all the attacks in the test set. 

\begin{figure*}[t]
       \centering
      \begin{subfigure}[t]{0.368\columnwidth}
        \includegraphics[width=\columnwidth]{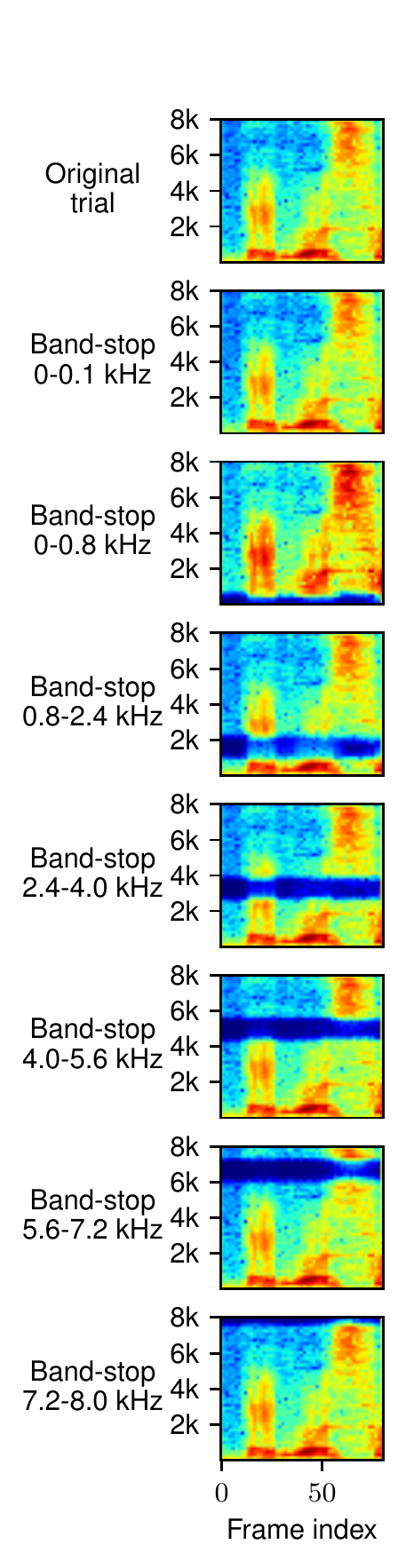}
        \label{fig:prob_sub1}
     \end{subfigure}
     \begin{subfigure}[t]{0.8\columnwidth}
        \includegraphics[width=\columnwidth]{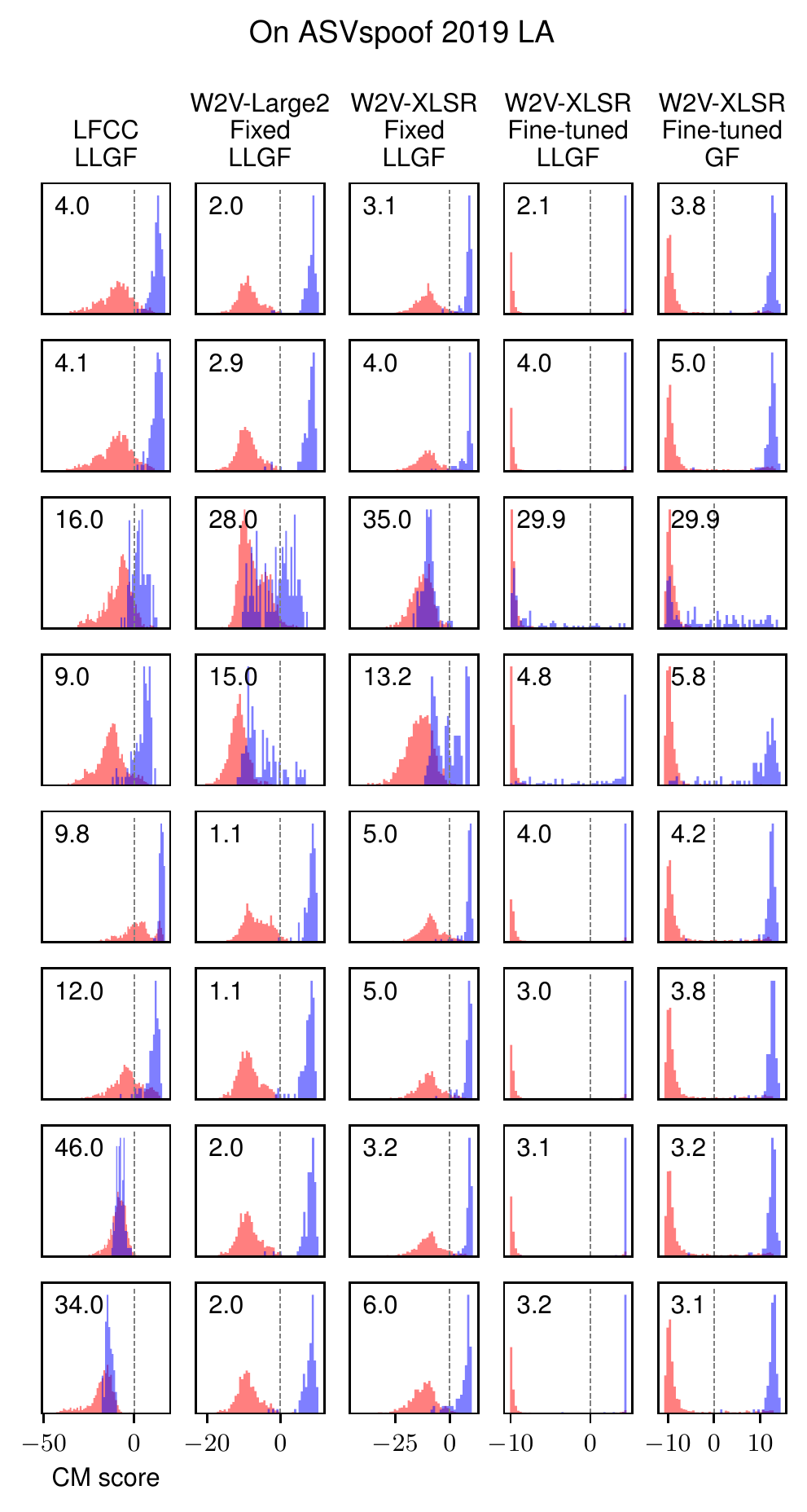}
        \label{fig:prob_sub2}
     \end{subfigure}
     \begin{subfigure}[t]{0.789\columnwidth}
        \includegraphics[width=\columnwidth]{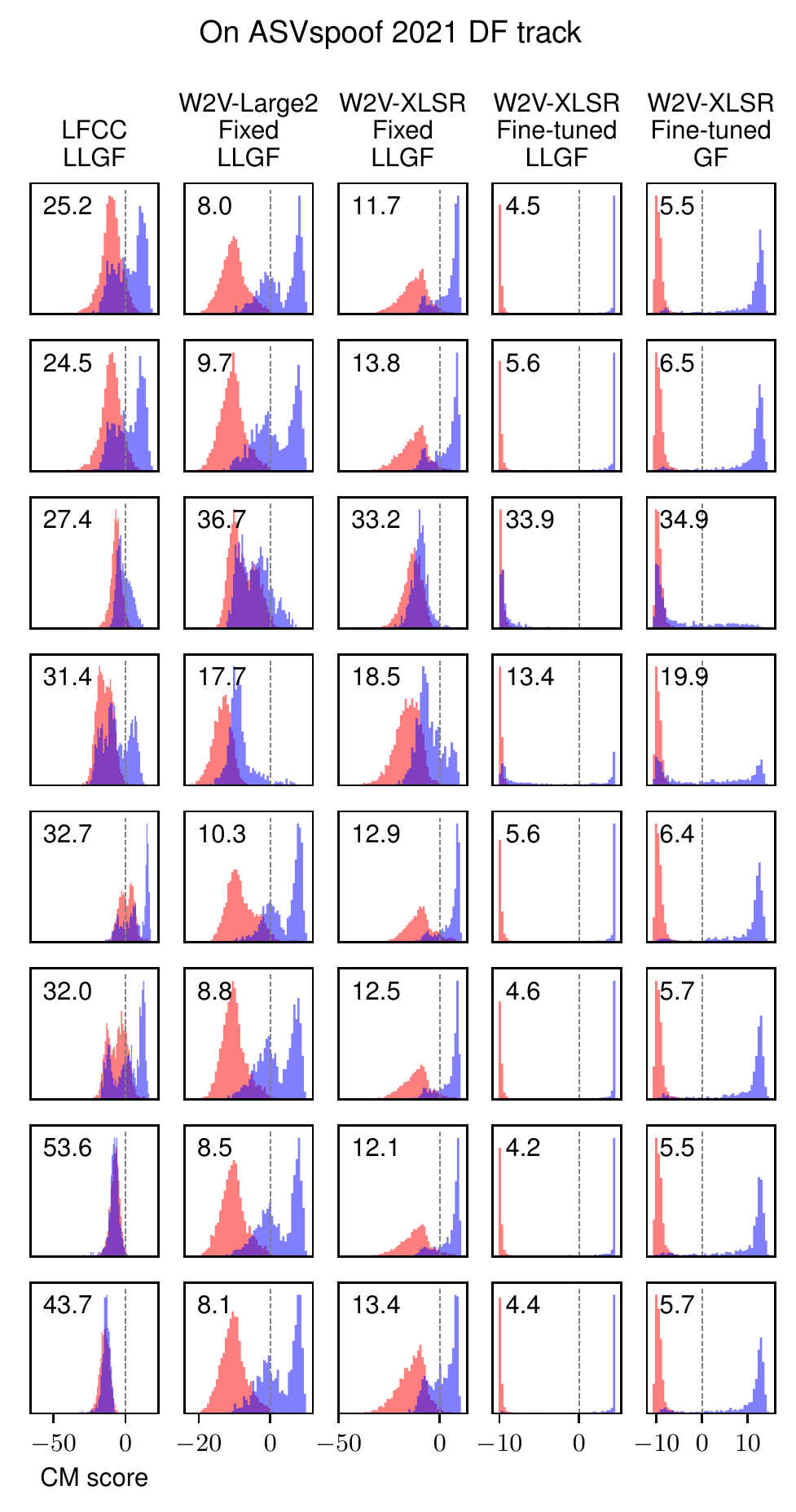}
        \label{fig:prob_sub3}
     \end{subfigure}
      \vspace{-6mm}
     \caption{Score distributions and EERs (\%) from band-stop probing test on subsets of ASVspoof 2019 LA and 2021 DF test sets. Score distributions of \textcolor{blue}{bona fide} and \textcolor{red}{spoofed} trials are in \textcolor{blue}{blue} and \textcolor{red}{red}, respectively. EER is shown in top-left corner of each score distribution sub-figure. Results in same row used same band-stop filtering configuration. Leftmost column plots spectrogram of  example trial in corresponding configuration. }
     \label{fig:prob}
\end{figure*}

\section{Sub-band analysis}
The results reported in our experiments motivated us to analyze the behavior of the CMs when a self-supervised front end is used. 
Recently, researchers have analyzed the features extracted from Wav2vec 2.0 \cite{Shah2021, Pasad2021}. 
They observed a correlation between the Wav2vec 2.0 features and certain acoustic (i.e., utterance-level zero crossing rate) or linguistic  (i.e., segmental-level phone identity) features. However, these works assume different down-stream tasks from anti-spoofing, and the outcomes do not address our question on anti-spoofing.

\subsection{Analysis configuration}

Among the many possible ways to probe CMs, we explored how a CM responds to band-stop filtered input trials. Our analysis is motivated by a similar sub-band analysis on CMs \cite{Tak2020}. Given the trials from a test set, we used a band-stop filter to filter out or mask a specified frequency band in the trials. After that, we fed the band-stop filtered trials to a well trained CM and obtained scores. By comparing the score distributions before and after the band-stop filtering, we can infer whether a CM is utilizing the information in a certain frequency band. 

The filters we used were 10th-order Butterworth filters, and the investigated stopbands were $0 - 0.1$, $0 - 0.8$, $0.8 - 2.4$,  $2.4 - 4.0$,  $4.0 - 5.6$,  $5.6 - 7.2$,  and $7.2 - 8.0$ kHz\footnote{In implementation, we used the SciPy 1.7.2 API \cite{Jones:2001aa} to filter the waveforms. For the first two stopbands, we used the high-pass filter API, and for the stopband of $7.2 - 8.0$ kHz, a low-pass filter was used. We refer to them as band-stop filtering for the explanation in this section.}. The spectrograms of a band-stop filtered trial are plotted on the left side of Figure~\ref{fig:prob}. The stopband at $0-0.1$ kHz was included to investigate the impact from the direct current offset, humming noise, and other components in this very low-frequency region. 

The analysis was conducted on five CMs built in the previous section: baseline using LFCC and \BNIIIshort, fixed \texttt{W2V-Large2} with \BNIshort, fixed \texttt{W2V-XLSR} with \BNIshort, fine-tuned \texttt{W2V-XLSR} with \BNIshort, and fine-tuned \texttt{W2V-XLSR} with \BNIIIshort. To reduce the computation time, we randomly selected trials from two test sets. One set consisted of 1,400 trials from the ASVspoof 2019 LA test set; the other contained 6,000 trials from the ASVspoof 2021 DF test set. The trial score distributions and the EER in each condition are plotted in Figure~\ref{fig:prob}. 

\subsection{Findings from sub-band analysis}
We first observed that the baseline CM using LFCC and \BNIIIshort\ was sensitive to the band-stop filtering in most of the sub-bands. In particular, as the last two rows of the figure show, the baseline CM's EERs degraded severely when the stopband was either $5.6-7.2$ or $7.2-8.0$ kHz. For example, when the band-stop filtering was on $5.6-7.2$ kHz, the EER on the ASVspoof 2019 LA test subset increased to 46\%. This indicates that the baseline CM trained on the ASVspoof 2019 LA training set relied heavily on the information in the high-frequency band. This trend was also reported in \cite{Tak2020}, even though the CM back end in that work was based on a Gaussian mixture model.

Different from the baseline, the four CMs using a self-supervised front end were relatively insensitive to the band-stop filtering on high-frequency bands. In fact, for the four CMs, the band-stop filtering at other sub-bands did not dramatically change the score distributions except for the two bands at $0 - 0.8$ and $0.8-2.4$ kHz.
However, when the band-stop filtering was conducted on either $0 - 0.8$ or $0.8-2.4$ kHz, the score distributions changed, and there was more overlap between those from the bona fide and spoofed trials. This trend was consistent across the four CMs and the two test subsets, except for the two CMs using fine-tuned \texttt{W2V-XLSR} on the ASVspoof 2019 test subset. There, using band-stop filtering at $0.8-2.4$ kHz only slightly increased the EERs to 4.8\% and 5.8\% for the two CMs, respectively.

Nevertheless, the results indicate that the CMs with a self-supervised front end mainly relied on the information between $0.1$ and $2.4$ kHz. This behavior is different from the baseline CM. One possible reason is that the self-supervised model was pre-trained on various speech data and tended to extract phone or high-level linguistic information while ignoring channel variation. The low-frequency band may be a good region to look for the desired linguistic information. Accordingly, the CM using a self-supervised front end tended to focus on that frequency band. 

If this explanation is true, we argue that the frequency band between 0.1 and 2.4 kHz contains useful information to discriminate bona fide and spoofed trials, and this information is generalizable to all the test sets in this study. This information helped the CMs using the fine-tuned \texttt{W2V-XLSR} achieve low EERs across all the test sets. The high-frequency band can be useful for a CM if it is only trained and tested on the ASVspoof 2019 LA data set. However, the discriminative information in the high-frequency band of the ASVspoof 2019 LA trials may not exist in the other datasets, which may have lead to the degraded performance on those datasets. 

\section{Conclusion}
\label{sec:con}
We investigated the use of self-supervised models as the front end of speech spoofing CMs. 
Through experiments on benchmark datasets, we observed that a self-supervised front end pre-trained using diverse speech data performed quite well when it is fixed and combined with a conventional LCNN-LSTM back end. 
More notable improvement is achieved when the front end is fine-tuned for the anti-spoofing task. Using only the 2019 LA training set, the CM with a fine-tuned front end not only performed decently on the 2019 LA test set but also significantly outperformed the baseline on the 2015, 2021 LA and 2021 DF test sets.
The results from a sub-band analysis further showed that CMs with a pre-trained self-supervised front end relied on information in the $0.1-2.4$ kHz frequency band to discriminate spoofed and bona fide trials, and this information is useful across test sets. 
Although the EERs reported in this study cannot be directly compared with other studies because the pre-trained self-supervised front end used more speech data, the results at least suggest one potential direction to improve the generalizability of CMs.

The self-supervised models can be used in different manners for down-stream tasks, e.g, partial re-randomization of pre-trained weights \cite{Pasad2021} and weighted sum of hidden features \cite{yang21c_interspeech}. We also tested some of the methods, even though they are not consistently better than the strategies discussed in this paper. These additional results are available in the Appendix \url{https://arxiv.org/abs/2111.07725}.

\bibliographystyle{IEEEbib}
\bibliography{library}

\appendix
\onecolumn

\section{Statistical analysis results}

\begin{figure*}[h]
\centering
\includegraphics[trim=0 200 0 100, clip, width=1.0\textwidth]{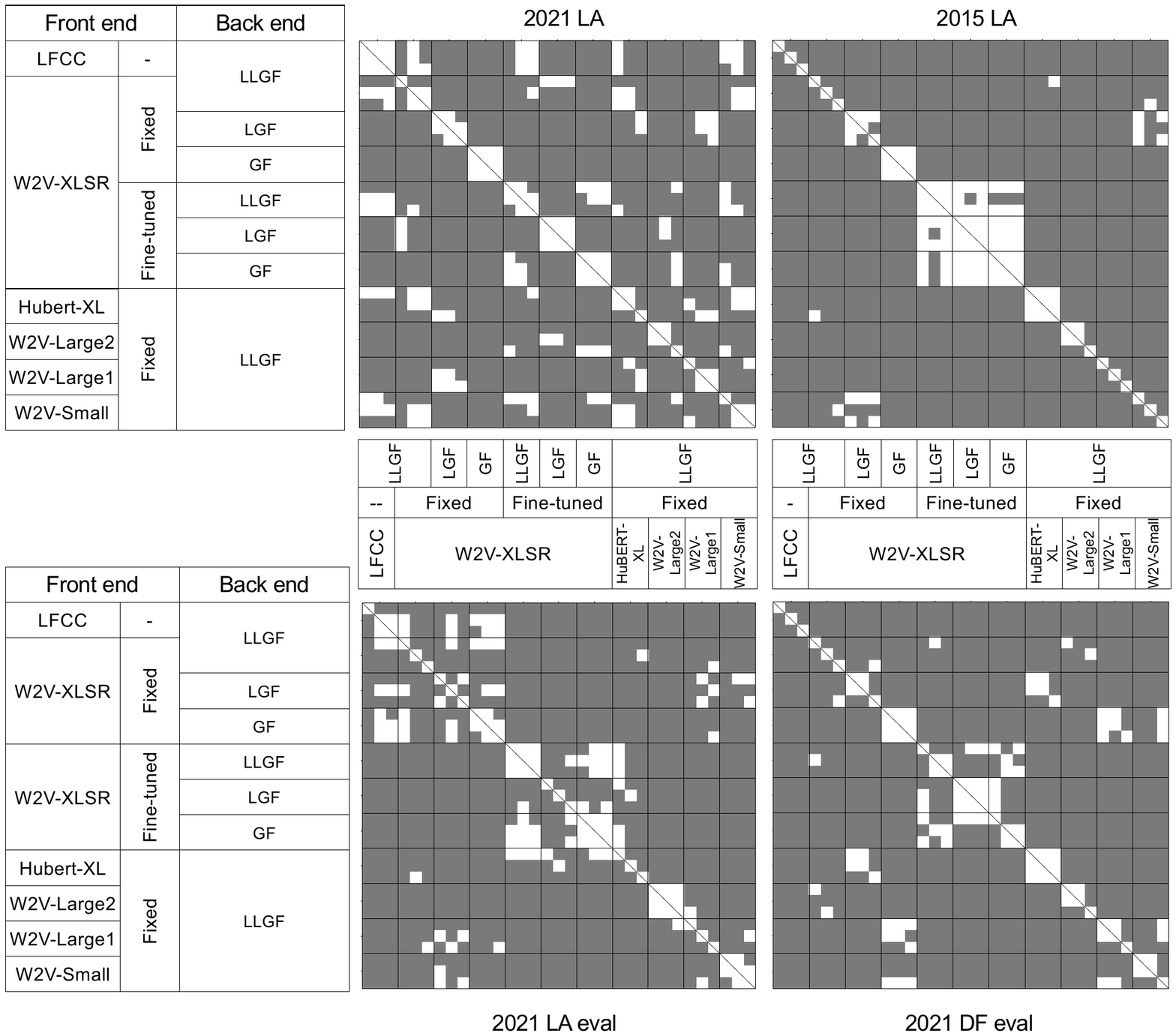}
\vspace{-8mm}
\caption{Statistical significance test using EERs on LA 2019 and Holm-Bonferroni correction with $\alpha=0.05$. Significant difference is indicated by dark gray, otherwise by white. Each square in the black frames contains $3\times3$ entries and denotes pair-wise tests between three training-evaluation rounds of two models. The three rounds of each model were in the same order as that in Table~\ref{tab:result}.}
\label{fig:sig_test}
\end{figure*}

\newpage
\section{Other experiment results}
\subsection{Min tDCF on 2021 LA eval track and 2019 LA test sets}

\begin{table*}[h]
\caption{Min tDCFs on different test sets. All the models were trained using the ASVspoof 2019 LA training set. A darker cell color indicates a higher EER value. The order of three rounds of each model follows that in Table~\ref{tab:result}. Min tDCF are computed following the official code in \url{https://github.com/asvspoof-challenge/2021}. Note that this min t-DCF for ASVspoof 2021 is slightly different from the legacy min t-DCF used in ASVspoof 2019. See more in the ASVspoof 2021 evaluation plan.}
\vspace{-5mm}
\begin{center}
\resizebox{\textwidth}{!}
{
\setlength{\tabcolsep}{4pt}
\begin{tabular}{l|ccc|ccc|ccc|ccc|ccc|ccc|ccc}
\toprule
Front end  &    \multicolumn{3}{c|}{LFCC} & \multicolumn{9}{c|}{\texttt{W2V-XLSR}, fixed} & \multicolumn{9}{c|}{\texttt{W2V-XLSR}, fine-tuned} \\
 \midrule
Back end &  \multicolumn{3}{c|}{\BNI} & \multicolumn{3}{c|}{\BNI} & \multicolumn{3}{c|}{\BNII} & \multicolumn{3}{c|}{\BNIII} & \multicolumn{3}{c|}{\BNI} & \multicolumn{3}{c|}{\BNII} & \multicolumn{3}{c|}{\BNIII} \\
 \midrule
              &   I   &  II   &  III  &   I   &  II   &  III  &   I   &  II   &  III  &   I   &  II   &  III  &   I   &  II   &  III  &   I   &  II   &  III  &   I   &  II   &  III \\ 
\midrule
 2019 LA    & \cellcolor[rgb]{0.98, 0.98, 0.98} 0.132 & \cellcolor[rgb]{0.99, 0.99, 0.99} 0.131 & \cellcolor[rgb]{0.98, 0.98, 0.98} 0.135 & \cellcolor[rgb]{0.99, 0.99, 0.99} 0.105 & \cellcolor[rgb]{0.98, 0.98, 0.98} 0.158 & \cellcolor[rgb]{0.97, 0.97, 0.97} 0.177 & \cellcolor[rgb]{0.96, 0.96, 0.96} 0.227 & \cellcolor[rgb]{0.96, 0.96, 0.96} 0.246 & \cellcolor[rgb]{0.95, 0.95, 0.95} 0.265 & \cellcolor[rgb]{0.87, 0.87, 0.87} 0.478 & \cellcolor[rgb]{0.86, 0.86, 0.86} 0.497 & \cellcolor[rgb]{0.86, 0.86, 0.86} 0.503 & \cellcolor[rgb]{0.99, 0.99, 0.99} 0.120 & \cellcolor[rgb]{0.99, 0.99, 0.99} 0.122 & \cellcolor[rgb]{0.99, 0.99, 0.99} 0.126 & \cellcolor[rgb]{0.99, 0.99, 0.99} 0.100 & \cellcolor[rgb]{0.99, 0.99, 0.99} 0.096 & \cellcolor[rgb]{0.99, 0.99, 0.99} 0.105 & \cellcolor[rgb]{0.99, 0.99, 0.99} 0.120 & \cellcolor[rgb]{0.99, 0.99, 0.99} 0.124 & \cellcolor[rgb]{0.99, 0.99, 0.99} 0.125\\
2021 LA eval. & \cellcolor[rgb]{0.76, 0.76, 0.76} 0.764 & \cellcolor[rgb]{0.77, 0.77, 0.77} 0.747 & \cellcolor[rgb]{0.78, 0.78, 0.78} 0.720 & \cellcolor[rgb]{0.91, 0.91, 0.91} 0.472 & \cellcolor[rgb]{0.81, 0.81, 0.81} 0.668 & \cellcolor[rgb]{0.83, 0.83, 0.83} 0.636 & \cellcolor[rgb]{0.77, 0.77, 0.77} 0.746 & \cellcolor[rgb]{0.83, 0.83, 0.83} 0.637 & \cellcolor[rgb]{0.83, 0.83, 0.83} 0.640 & \cellcolor[rgb]{0.82, 0.82, 0.82} 0.648 & \cellcolor[rgb]{0.81, 0.81, 0.81} 0.667 & \cellcolor[rgb]{0.81, 0.81, 0.81} 0.674 & \cellcolor[rgb]{0.95, 0.95, 0.95} 0.377 & \cellcolor[rgb]{0.95, 0.95, 0.95} 0.362 & \cellcolor[rgb]{0.95, 0.95, 0.95} 0.359 & \cellcolor[rgb]{0.95, 0.95, 0.95} 0.378 & \cellcolor[rgb]{0.95, 0.95, 0.95} 0.348 & \cellcolor[rgb]{0.96, 0.96, 0.96} 0.340 & \cellcolor[rgb]{0.95, 0.95, 0.95} 0.373 & \cellcolor[rgb]{0.94, 0.94, 0.94} 0.392 & \cellcolor[rgb]{0.95, 0.95, 0.95} 0.364\\ 
\bottomrule
\multicolumn{4}{c}{} & \multicolumn{3}{c}{($\Downarrow$ results are copied)} & \multicolumn{12}{c}{} & $\phantom{00.00}$ & $\phantom{00.00}$ & $\phantom{00.00}$  \\
\cmidrule{1-16}
Front end  &   \multicolumn{3}{c|}{\texttt{HuBERT-XL}, fixed} &     \multicolumn{3}{c|}{\texttt{W2V-XLSR}, fixed} &    \multicolumn{3}{c|}{\texttt{W2V-Large2}, fixed}  &    \multicolumn{3}{c|}{\texttt{W2V-Large1}, fixed} &    \multicolumn{3}{c|}{\texttt{W2V-Small}, fixed} &    \multicolumn{3}{c}{}  \\
\cmidrule{1-16}
Back end &  \multicolumn{15}{c|}{\BNI} & \multicolumn{3}{c}{} \\
\cmidrule{1-16}
   &   I   &  II   &  III  &   I   &  II   &  III  &   I   &  II   &  III  &   I   &  II   &  III &   I   &  II   &  III & \multicolumn{3}{c}{} \\
\cmidrule{1-16}
 2019 LA    & \cellcolor[rgb]{0.98, 0.98, 0.98} 0.157 & \cellcolor[rgb]{0.98, 0.98, 0.98} 0.164 & \cellcolor[rgb]{0.96, 0.96, 0.96} 0.217 & \cellcolor[rgb]{0.99, 0.99, 0.99} 0.105 & \cellcolor[rgb]{0.98, 0.98, 0.98} 0.158 & \cellcolor[rgb]{0.97, 0.97, 0.97} 0.177 & \cellcolor[rgb]{1.00, 1.00, 1.00} 0.088 & \cellcolor[rgb]{0.99, 0.99, 0.99} 0.093 & \cellcolor[rgb]{0.99, 0.99, 0.99} 0.125 & \cellcolor[rgb]{0.97, 0.97, 0.97} 0.204 & \cellcolor[rgb]{0.96, 0.96, 0.96} 0.234 & \cellcolor[rgb]{0.95, 0.95, 0.95} 0.263 & \cellcolor[rgb]{0.98, 0.98, 0.98} 0.139 & \cellcolor[rgb]{0.97, 0.97, 0.97} 0.173 & \cellcolor[rgb]{0.97, 0.97, 0.97} 0.185\\ 
2021 LA eval. & \cellcolor[rgb]{0.92, 0.92, 0.92} 0.430 & \cellcolor[rgb]{0.95, 0.95, 0.95} 0.354 & \cellcolor[rgb]{0.92, 0.92, 0.92} 0.437 & \cellcolor[rgb]{0.91, 0.91, 0.91} 0.472 & \cellcolor[rgb]{0.81, 0.81, 0.81} 0.668 & \cellcolor[rgb]{0.83, 0.83, 0.83} 0.636 & \cellcolor[rgb]{0.88, 0.88, 0.88} 0.521 & \cellcolor[rgb]{0.89, 0.89, 0.89} 0.506 & \cellcolor[rgb]{0.88, 0.88, 0.88} 0.531 & \cellcolor[rgb]{0.86, 0.86, 0.86} 0.568 & \cellcolor[rgb]{0.82, 0.82, 0.82} 0.656 & \cellcolor[rgb]{0.79, 0.79, 0.79} 0.713 & \cellcolor[rgb]{0.83, 0.83, 0.83} 0.639 & \cellcolor[rgb]{0.86, 0.86, 0.86} 0.573 & \cellcolor[rgb]{0.85, 0.85, 0.85} 0.598\\ 
\cmidrule{1-16}
\end{tabular}
}
\end{center}
\vspace{-5mm}
\label{tab:result2}
\end{table*}

\subsection{Decomposed EERs on ASVspoof 2021 based on attacker type}

\begin{table*}[h]
\caption{Decomposed EERs (\%) on the test sets. EERs from the three training-evaluation rounds are averaged for each model.}
\vspace{-5mm}
\begin{center}
\resizebox{0.7\textwidth}{!}
{
\setlength{\tabcolsep}{2pt}
\begin{tabular}{cccccccccccccccccc}
\toprule
\multicolumn{3}{c}{CM config}  & &  \multicolumn{13}{c}{ASVspoof 2021 LA test set (2021 LA)}  \\  
\cmidrule(lr){1-3}\cmidrule(lr){5-17}
\multicolumn{2}{c}{\multirow{2}{*}{Front end}} & \multirow{2}{*}{{Back end}}& & \multicolumn{2}{c}{Known attack} & \multicolumn{4}{c}{Partially known attack} & \multicolumn{7}{c}{Unknown attack} \\
\cmidrule(lr){5-6}\cmidrule(lr){7-10}\cmidrule(lr){11-17}
& &  &  &  A16  &  A19  &  A07  &  A08  &  A09  &  A17  &  A10  &  A11  &  A12  &  A13  &  A14  &  A15  &  A18 \\  
\midrule
LFCC & -- & \BNIshort\                                                       &   & \cellcolor[rgb]{0.72, 0.72, 0.72} 20.85 & \cellcolor[rgb]{0.61, 0.61, 0.61} 38.48 & \cellcolor[rgb]{0.72, 0.72, 0.72} 21.05 & \cellcolor[rgb]{0.78, 0.78, 0.78} 14.57 & \cellcolor[rgb]{0.82, 0.82, 0.82} 10.89 & \cellcolor[rgb]{0.60, 0.60, 0.60} 40.89 & \cellcolor[rgb]{0.70, 0.70, 0.70} 23.00 & \cellcolor[rgb]{0.76, 0.76, 0.76} 16.11 & \cellcolor[rgb]{0.72, 0.72, 0.72} 21.54 & \cellcolor[rgb]{0.70, 0.70, 0.70} 22.94 & \cellcolor[rgb]{0.75, 0.75, 0.75} 17.75 & \cellcolor[rgb]{0.74, 0.74, 0.74} 19.03 & \cellcolor[rgb]{0.63, 0.63, 0.63} 33.66\\ 
\midrule
  \multirow{5}{*}{\texttt{W2V-XLSR}} & \multirow{3}{*}{{Fixed}} & \BNIshort\  && \cellcolor[rgb]{0.78, 0.78, 0.78} 14.13 & \cellcolor[rgb]{0.77, 0.77, 0.77} 14.84 & \cellcolor[rgb]{0.75, 0.75, 0.75} 18.35 & \cellcolor[rgb]{0.73, 0.73, 0.73} 19.52 & \cellcolor[rgb]{0.80, 0.80, 0.80} 12.71 & \cellcolor[rgb]{0.87, 0.87, 0.87} 6.77 & \cellcolor[rgb]{0.66, 0.66, 0.66} 28.90 & \cellcolor[rgb]{0.75, 0.75, 0.75} 17.96 & \cellcolor[rgb]{0.76, 0.76, 0.76} 16.35 & \cellcolor[rgb]{0.87, 0.87, 0.87} 6.38 & \cellcolor[rgb]{0.78, 0.78, 0.78} 14.25 & \cellcolor[rgb]{0.74, 0.74, 0.74} 18.54 & \cellcolor[rgb]{0.83, 0.83, 0.83} 9.84\\ 
 & & \BNIIshort\ &                                                             & \cellcolor[rgb]{0.76, 0.76, 0.76} 16.32 & \cellcolor[rgb]{0.71, 0.71, 0.71} 22.12 & \cellcolor[rgb]{0.77, 0.77, 0.77} 15.53 & \cellcolor[rgb]{0.76, 0.76, 0.76} 16.87 & \cellcolor[rgb]{0.77, 0.77, 0.77} 15.43 & \cellcolor[rgb]{0.81, 0.81, 0.81} 11.33 & \cellcolor[rgb]{0.71, 0.71, 0.71} 22.40 & \cellcolor[rgb]{0.72, 0.72, 0.72} 20.69 & \cellcolor[rgb]{0.76, 0.76, 0.76} 16.05 & \cellcolor[rgb]{0.79, 0.79, 0.79} 13.38 & \cellcolor[rgb]{0.77, 0.77, 0.77} 15.46 & \cellcolor[rgb]{0.72, 0.72, 0.72} 21.09 & \cellcolor[rgb]{0.80, 0.80, 0.80} 12.19\\ 
 & & \BNIIIshort\ &                                                            & \cellcolor[rgb]{0.73, 0.73, 0.73} 19.80 & \cellcolor[rgb]{0.61, 0.61, 0.61} 37.96 & \cellcolor[rgb]{0.80, 0.80, 0.80} 12.04 & \cellcolor[rgb]{0.74, 0.74, 0.74} 18.81 & \cellcolor[rgb]{0.75, 0.75, 0.75} 17.23 & \cellcolor[rgb]{0.72, 0.72, 0.72} 21.34 & \cellcolor[rgb]{0.76, 0.76, 0.76} 16.18 & \cellcolor[rgb]{0.80, 0.80, 0.80} 12.62 & \cellcolor[rgb]{0.77, 0.77, 0.77} 15.84 & \cellcolor[rgb]{0.91, 0.91, 0.91} 3.52 & \cellcolor[rgb]{0.67, 0.67, 0.67} 27.11 & \cellcolor[rgb]{0.61, 0.61, 0.61} 36.81 & \cellcolor[rgb]{0.71, 0.71, 0.71} 21.95\\ 
\cmidrule{2-17}
 & \multirow{3}{*}{\shortstack{Fine- \\ tuned}} & \BNIshort\  &                & \cellcolor[rgb]{0.93, 0.93, 0.93} 2.62 & \cellcolor[rgb]{0.92, 0.92, 0.92} 3.24 & \cellcolor[rgb]{0.91, 0.91, 0.91} 3.66 & \cellcolor[rgb]{0.90, 0.90, 0.90} 4.05 & \cellcolor[rgb]{0.92, 0.92, 0.92} 3.15 & \cellcolor[rgb]{0.93, 0.93, 0.93} 2.75 & \cellcolor[rgb]{0.78, 0.78, 0.78} 14.72 & \cellcolor[rgb]{0.75, 0.75, 0.75} 18.00 & \cellcolor[rgb]{0.92, 0.92, 0.92} 3.21 & \cellcolor[rgb]{0.93, 0.93, 0.93} 2.76 & \cellcolor[rgb]{0.88, 0.88, 0.88} 5.93 & \cellcolor[rgb]{0.86, 0.86, 0.86} 7.46 & \cellcolor[rgb]{0.92, 0.92, 0.92} 3.16\\ 
 & &  \BNIIshort\ &                                                            & \cellcolor[rgb]{0.94, 0.94, 0.94} 2.08 & \cellcolor[rgb]{0.92, 0.92, 0.92} 2.99 & \cellcolor[rgb]{0.93, 0.93, 0.93} 2.61 & \cellcolor[rgb]{0.92, 0.92, 0.92} 3.05 & \cellcolor[rgb]{0.94, 0.94, 0.94} 2.26 & \cellcolor[rgb]{0.94, 0.94, 0.94} 2.47 & \cellcolor[rgb]{0.75, 0.75, 0.75} 17.42 & \cellcolor[rgb]{0.73, 0.73, 0.73} 19.57 & \cellcolor[rgb]{0.93, 0.93, 0.93} 2.77 & \cellcolor[rgb]{0.96, 0.96, 0.96} 1.10 & \cellcolor[rgb]{0.89, 0.89, 0.89} 4.83 & \cellcolor[rgb]{0.84, 0.84, 0.84} 9.11 & \cellcolor[rgb]{0.92, 0.92, 0.92} 3.11\\ 
 & &  \BNIIIshort\ &                                                           & \cellcolor[rgb]{0.94, 0.94, 0.94} 2.37 & \cellcolor[rgb]{0.91, 0.91, 0.91} 3.73 & \cellcolor[rgb]{0.90, 0.90, 0.90} 4.16 & \cellcolor[rgb]{0.91, 0.91, 0.91} 3.84 & \cellcolor[rgb]{0.94, 0.94, 0.94} 2.29 & \cellcolor[rgb]{0.92, 0.92, 0.92} 3.26 & \cellcolor[rgb]{0.74, 0.74, 0.74} 18.54 & \cellcolor[rgb]{0.74, 0.74, 0.74} 18.82 & \cellcolor[rgb]{0.93, 0.93, 0.93} 2.78 & \cellcolor[rgb]{0.93, 0.93, 0.93} 2.83 & \cellcolor[rgb]{0.88, 0.88, 0.88} 5.54 & \cellcolor[rgb]{0.85, 0.85, 0.85} 7.83 & \cellcolor[rgb]{0.92, 0.92, 0.92} 3.33\\ 
 \midrule
 \texttt{HuBERT-XL} & \multirow{4}{*}{{Fixed}}  & \multirow{4}{*}{{\BNIshort}}&& \cellcolor[rgb]{0.88, 0.88, 0.88} 5.98 & \cellcolor[rgb]{0.76, 0.76, 0.76} 16.39 & \cellcolor[rgb]{0.87, 0.87, 0.87} 6.19 & \cellcolor[rgb]{0.85, 0.85, 0.85} 8.29 & \cellcolor[rgb]{0.92, 0.92, 0.92} 3.22 & \cellcolor[rgb]{0.89, 0.89, 0.89} 5.13 & \cellcolor[rgb]{0.79, 0.79, 0.79} 13.65 & \cellcolor[rgb]{0.86, 0.86, 0.86} 7.35 & \cellcolor[rgb]{0.92, 0.92, 0.92} 3.40 & \cellcolor[rgb]{0.97, 0.97, 0.97} 0.69 & \cellcolor[rgb]{0.85, 0.85, 0.85} 8.41 & \cellcolor[rgb]{0.86, 0.86, 0.86} 7.62 & \cellcolor[rgb]{0.76, 0.76, 0.76} 16.63\\ 
 \texttt{W2V-Large2} & &  &                                                    & \cellcolor[rgb]{0.82, 0.82, 0.82} 10.71 & \cellcolor[rgb]{0.81, 0.81, 0.81} 11.76 & \cellcolor[rgb]{0.78, 0.78, 0.78} 14.30 & \cellcolor[rgb]{0.80, 0.80, 0.80} 12.43 & \cellcolor[rgb]{0.88, 0.88, 0.88} 5.69 & \cellcolor[rgb]{0.86, 0.86, 0.86} 7.29 & \cellcolor[rgb]{0.72, 0.72, 0.72} 21.14 & \cellcolor[rgb]{0.81, 0.81, 0.81} 11.92 & \cellcolor[rgb]{0.80, 0.80, 0.80} 12.63 & \cellcolor[rgb]{0.91, 0.91, 0.91} 3.69 & \cellcolor[rgb]{0.80, 0.80, 0.80} 12.69 & \cellcolor[rgb]{0.77, 0.77, 0.77} 15.44 & \cellcolor[rgb]{0.88, 0.88, 0.88} 5.39\\ 
 \texttt{W2V-Large1} & &  &                                                    & \cellcolor[rgb]{0.78, 0.78, 0.78} 14.20 & \cellcolor[rgb]{0.75, 0.75, 0.75} 17.65 & \cellcolor[rgb]{0.75, 0.75, 0.75} 17.87 & \cellcolor[rgb]{0.77, 0.77, 0.77} 15.08 & \cellcolor[rgb]{0.82, 0.82, 0.82} 10.20 & \cellcolor[rgb]{0.77, 0.77, 0.77} 15.77 & \cellcolor[rgb]{0.72, 0.72, 0.72} 21.79 & \cellcolor[rgb]{0.78, 0.78, 0.78} 14.58 & \cellcolor[rgb]{0.79, 0.79, 0.79} 13.01 & \cellcolor[rgb]{0.82, 0.82, 0.82} 10.89 & \cellcolor[rgb]{0.77, 0.77, 0.77} 15.32 & \cellcolor[rgb]{0.75, 0.75, 0.75} 17.36 & \cellcolor[rgb]{0.77, 0.77, 0.77} 15.78\\ 
 \texttt{W2V-Small} & &  &                                                     & \cellcolor[rgb]{0.78, 0.78, 0.78} 13.74 & \cellcolor[rgb]{0.76, 0.76, 0.76} 16.33 & \cellcolor[rgb]{0.74, 0.74, 0.74} 18.83 & \cellcolor[rgb]{0.82, 0.82, 0.82} 10.72 & \cellcolor[rgb]{0.89, 0.89, 0.89} 5.27 & \cellcolor[rgb]{0.76, 0.76, 0.76} 17.13 & \cellcolor[rgb]{0.71, 0.71, 0.71} 21.89 & \cellcolor[rgb]{0.84, 0.84, 0.84} 9.13 & \cellcolor[rgb]{0.82, 0.82, 0.82} 10.77 & \cellcolor[rgb]{0.86, 0.86, 0.86} 7.06 & \cellcolor[rgb]{0.79, 0.79, 0.79} 13.03 & \cellcolor[rgb]{0.76, 0.76, 0.76} 15.96 & \cellcolor[rgb]{0.77, 0.77, 0.77} 15.50\\
\bottomrule
\end{tabular}
}
\end{center}
\vspace{-5mm}
\label{fig:sig_alternative}
\end{table*}

\newpage
\subsection{Other experimental models}
To keep the main text concise, we do not list all the experimental results there. In this section, we show the results from other experimental models.

\subsubsection{Fine-tune on \texttt{W2V-Large2}}
In Table~\ref{tab:result}, only \texttt{W2V-XLSR} was fine-tuned. Here, we compare the \texttt{W2V-XLSR} and \texttt{W2V-Large2} in the fined-tuned condition. The EERs are not so different. Both \texttt{W2V-XLSR} and \texttt{W2V-Large2} are reasonably good the antis-spoofing task if we want to fine tune the model. 

\begin{table*}[h]
\caption{EERs (\%) on different test sets. All the models were trained using the ASVspoof 2019 LA training set. A darker cell color indicates a higher EER value. Columns with \textbf{a title in bold font are new results}.}
\vspace{-5mm}
\begin{center}
\resizebox{\textwidth}{!}
{
\setlength{\tabcolsep}{4pt}
\begin{tabular}{l|ccc|ccc|ccc|ccc|ccc|ccc}
\toprule
Front end  &    \multicolumn{9}{c|}{\texttt{W2V-XLSR}, fine-tuned (EERs copied from Table~\ref{tab:result})}  &    \multicolumn{9}{c|}{\texttt{\textbf{W2V-Large2}}, \textbf{fine-tuned}}  \\
 \midrule
Back end &  \multicolumn{3}{c|}{\BNI} & \multicolumn{3}{c|}{\BNII} & \multicolumn{3}{c|}{\BNIII} & \multicolumn{3}{c|}{\BNI} & \multicolumn{3}{c|}{\BNII} & \multicolumn{3}{c|}{\BNIII} \\
 \midrule
              &   I   &  II   &  III  &   I   &  II   &  III  &   I   &  II   &  III  &   I   &  II   &  III  &   I   &  II   &  III  &   I   &  II   &  III  \\ 
\midrule
   2019 LA    & \cellcolor[rgb]{0.96, 0.96, 0.96} 2.31 & \cellcolor[rgb]{0.95, 0.95, 0.95} 2.80 & \cellcolor[rgb]{0.94, 0.94, 0.94} 3.08 & \cellcolor[rgb]{0.98, 0.98, 0.98} 1.28 & \cellcolor[rgb]{0.98, 0.98, 0.98} 1.28 & \cellcolor[rgb]{0.97, 0.97, 0.97} 1.50 & \cellcolor[rgb]{0.96, 0.96, 0.96} 1.96 & \cellcolor[rgb]{0.96, 0.96, 0.96} 2.25 & \cellcolor[rgb]{0.96, 0.96, 0.96} 2.27 & \cellcolor[rgb]{0.97, 0.97, 0.97} 1.86 & \cellcolor[rgb]{0.96, 0.96, 0.96} 2.42 & \cellcolor[rgb]{0.94, 0.94, 0.94} 2.99 & \cellcolor[rgb]{0.98, 0.98, 0.98} 1.13 & \cellcolor[rgb]{0.98, 0.98, 0.98} 1.28 & \cellcolor[rgb]{0.91, 0.91, 0.91} 4.14 & \cellcolor[rgb]{0.98, 0.98, 0.98} 1.29 & \cellcolor[rgb]{0.96, 0.96, 0.96} 2.01 & \cellcolor[rgb]{0.93, 0.93, 0.93} 3.51\\ 
   2015 LA    & \cellcolor[rgb]{1.00, 1.00, 1.00} 0.25 & \cellcolor[rgb]{1.00, 1.00, 1.00} 0.41 & \cellcolor[rgb]{1.00, 1.00, 1.00} 0.24 & \cellcolor[rgb]{1.00, 1.00, 1.00} 0.24 & \cellcolor[rgb]{1.00, 1.00, 1.00} 0.19 & \cellcolor[rgb]{1.00, 1.00, 1.00} 0.31 & \cellcolor[rgb]{1.00, 1.00, 1.00} 0.21 & \cellcolor[rgb]{1.00, 1.00, 1.00} 0.17 & \cellcolor[rgb]{1.00, 1.00, 1.00} 0.17 & \cellcolor[rgb]{1.00, 1.00, 1.00} 0.32 & \cellcolor[rgb]{1.00, 1.00, 1.00} 0.32 & \cellcolor[rgb]{1.00, 1.00, 1.00} 0.34 & \cellcolor[rgb]{1.00, 1.00, 1.00} 0.25 & \cellcolor[rgb]{1.00, 1.00, 1.00} 0.16 & \cellcolor[rgb]{1.00, 1.00, 1.00} 0.26 & \cellcolor[rgb]{1.00, 1.00, 1.00} 0.33 & \cellcolor[rgb]{1.00, 1.00, 1.00} 0.30 & \cellcolor[rgb]{1.00, 1.00, 1.00} 0.30\\ 
2021 LA prog. & \cellcolor[rgb]{0.80, 0.80, 0.80} 7.58 & \cellcolor[rgb]{0.84, 0.84, 0.84} 6.38 & \cellcolor[rgb]{0.83, 0.83, 0.83} 6.56 & \cellcolor[rgb]{0.66, 0.66, 0.66} 10.63 & \cellcolor[rgb]{0.73, 0.73, 0.73} 9.19 & \cellcolor[rgb]{0.85, 0.85, 0.85} 6.27 & \cellcolor[rgb]{0.80, 0.80, 0.80} 7.65 & \cellcolor[rgb]{0.81, 0.81, 0.81} 7.16 & \cellcolor[rgb]{0.79, 0.79, 0.79} 7.82 & \cellcolor[rgb]{0.80, 0.80, 0.80} 7.52 & \cellcolor[rgb]{0.80, 0.80, 0.80} 7.64 & \cellcolor[rgb]{0.72, 0.72, 0.72} 9.43 & \cellcolor[rgb]{0.91, 0.91, 0.91} 4.06 & \cellcolor[rgb]{0.79, 0.79, 0.79} 7.70 & \cellcolor[rgb]{0.81, 0.81, 0.81} 7.28 & \cellcolor[rgb]{0.88, 0.88, 0.88} 5.31 & \cellcolor[rgb]{0.86, 0.86, 0.86} 5.79 & \cellcolor[rgb]{0.79, 0.79, 0.79} 7.70\\ 
2021 LA eval. & \cellcolor[rgb]{0.80, 0.80, 0.80} 7.62 & \cellcolor[rgb]{0.81, 0.81, 0.81} 7.26 & \cellcolor[rgb]{0.81, 0.81, 0.81} 7.18 & \cellcolor[rgb]{0.71, 0.71, 0.71} 9.66 & \cellcolor[rgb]{0.78, 0.78, 0.78} 8.11 & \cellcolor[rgb]{0.84, 0.84, 0.84} 6.53 & \cellcolor[rgb]{0.78, 0.78, 0.78} 7.99 & \cellcolor[rgb]{0.80, 0.80, 0.80} 7.42 & \cellcolor[rgb]{0.80, 0.80, 0.80} 7.61 & \cellcolor[rgb]{0.78, 0.78, 0.78} 7.98 & \cellcolor[rgb]{0.78, 0.78, 0.78} 8.03 & \cellcolor[rgb]{0.71, 0.71, 0.71} 9.74 & \cellcolor[rgb]{0.87, 0.87, 0.87} 5.51 & \cellcolor[rgb]{0.80, 0.80, 0.80} 7.57 & \cellcolor[rgb]{0.80, 0.80, 0.80} 7.39 & \cellcolor[rgb]{0.80, 0.80, 0.80} 7.55 & \cellcolor[rgb]{0.85, 0.85, 0.85} 6.10 & \cellcolor[rgb]{0.79, 0.79, 0.79} 7.80\\ 
2021 DF prog. & \cellcolor[rgb]{0.90, 0.90, 0.90} 4.40 & \cellcolor[rgb]{0.90, 0.90, 0.90} 4.33 & \cellcolor[rgb]{0.91, 0.91, 0.91} 4.14 & \cellcolor[rgb]{0.94, 0.94, 0.94} 3.38 & \cellcolor[rgb]{0.92, 0.92, 0.92} 3.75 & \cellcolor[rgb]{0.93, 0.93, 0.93} 3.55 & \cellcolor[rgb]{0.92, 0.92, 0.92} 3.97 & \cellcolor[rgb]{0.91, 0.91, 0.91} 4.23 & \cellcolor[rgb]{0.89, 0.89, 0.89} 4.94 & \cellcolor[rgb]{0.91, 0.91, 0.91} 4.13 & \cellcolor[rgb]{0.91, 0.91, 0.91} 4.23 & \cellcolor[rgb]{0.88, 0.88, 0.88} 5.08 & \cellcolor[rgb]{0.96, 0.96, 0.96} 2.29 & \cellcolor[rgb]{0.96, 0.96, 0.96} 2.46 & \cellcolor[rgb]{0.93, 0.93, 0.93} 3.68 & \cellcolor[rgb]{0.94, 0.94, 0.94} 3.30 & \cellcolor[rgb]{0.93, 0.93, 0.93} 3.47 & \cellcolor[rgb]{0.90, 0.90, 0.90} 4.59\\ 
2021 DF eval. & \cellcolor[rgb]{0.87, 0.87, 0.87} 5.44 & \cellcolor[rgb]{0.83, 0.83, 0.83} 6.68 & \cellcolor[rgb]{0.85, 0.85, 0.85} 6.18 & \cellcolor[rgb]{0.89, 0.89, 0.89} 4.75 & \cellcolor[rgb]{0.88, 0.88, 0.88} 5.23 & \cellcolor[rgb]{0.88, 0.88, 0.88} 4.98 & \cellcolor[rgb]{0.88, 0.88, 0.88} 5.04 & \cellcolor[rgb]{0.85, 0.85, 0.85} 6.10 & \cellcolor[rgb]{0.86, 0.86, 0.86} 5.88 & \cellcolor[rgb]{0.88, 0.88, 0.88} 5.23 & \cellcolor[rgb]{0.87, 0.87, 0.87} 5.56 & \cellcolor[rgb]{0.87, 0.87, 0.87} 5.36 & \cellcolor[rgb]{0.86, 0.86, 0.86} 5.62 & \cellcolor[rgb]{0.85, 0.85, 0.85} 6.07 & \cellcolor[rgb]{0.85, 0.85, 0.85} 6.08 & \cellcolor[rgb]{0.87, 0.87, 0.87} 5.35 & \cellcolor[rgb]{0.86, 0.86, 0.86} 5.87 & \cellcolor[rgb]{0.89, 0.89, 0.89} 4.95\\ 
\bottomrule
\end{tabular}
}
\end{center}
\vspace{-5mm}
\label{app-tab:result}
\end{table*}

\subsubsection{Partially re-randomizing pre-trained SSL}
When we use a pre-trained SSL, either fixing or fine-tuning it, we use all the weights from the pre-trained SSLs. However, in the case of fine-tuning a pre-trained SSL, we may re-randomize some of the weights before `fine-tuning'. In other words, we don't use all the pre-trained model weights. This strategy was proposed by Pasad et. al. \cite{Pasad2021}. They observed that the last a few Transformer blocks of the Wav2vec 2.0 model behave quite differently from other blocks. Therefore, they argue that it can be beneficial to discard the pre-trained weights from those blocks.

We can use this strategy in our task and re-randomized the last 3 Transformer blocks.  These layers produced hidden features that look quite different from other blocks. Table below shows the results.
We can see that the EER on 2019 LA test set were reduced. However,  the improvement is not always obvious on other test sets.  For example, when using a \BNI\ back end, the EERs on ASVspoof 2021 LA eval may slightly increase (e.g., 8.21\%) if the new strategy was used. 

This training strategy may be tried if the reader is interested in it. Note that the number of layers to be re-randomized has to be decided after analyzing the hidden features from the wav2vec 2.0 models, which requires additional efforts.

\begin{table*}[h]
\caption{EERs (\%) on different test sets. All the models were trained using the ASVspoof 2019 LA training set. A darker cell color indicates a higher EER value. Columns with \textbf{a title in bold font are new results}.}
\vspace{-5mm}
\begin{center}
\resizebox{\textwidth}{!}
{
\setlength{\tabcolsep}{4pt}
\begin{tabular}{l|ccc|ccc|ccc|ccc|ccc|ccc}
\toprule
Front end  &    \multicolumn{9}{c|}{\texttt{W2V-XLSR}, fine-tuned (EERs copied from Table~\ref{tab:result})}  &    \multicolumn{9}{c|}{\textbf{\texttt{W2V-XLSR}, fine-tuned, re-randomized last 3 Trans. blocks.}}  \\
 \midrule
Back end &  \multicolumn{3}{c|}{\BNI} & \multicolumn{3}{c|}{\BNII} & \multicolumn{3}{c|}{\BNIII} & \multicolumn{3}{c|}{\BNI} & \multicolumn{3}{c|}{\BNII} & \multicolumn{3}{c|}{\BNIII} \\
 \midrule
              &   I   &  II   &  III  &   I   &  II   &  III  &   I   &  II   &  III  &   I   &  II   &  III  &   I   &  II   &  III  &   I   &  II   &  III  \\ 
\midrule
   2019 LA    & \cellcolor[rgb]{0.96, 0.96, 0.96} 2.31 & \cellcolor[rgb]{0.95, 0.95, 0.95} 2.80 & \cellcolor[rgb]{0.94, 0.94, 0.94} 3.08 & \cellcolor[rgb]{0.98, 0.98, 0.98} 1.28 & \cellcolor[rgb]{0.98, 0.98, 0.98} 1.28 & \cellcolor[rgb]{0.97, 0.97, 0.97} 1.50 & \cellcolor[rgb]{0.96, 0.96, 0.96} 1.96 & \cellcolor[rgb]{0.96, 0.96, 0.96} 2.25 & \cellcolor[rgb]{0.96, 0.96, 0.96} 2.27 & \cellcolor[rgb]{0.97, 0.97, 0.97} 1.86 & \cellcolor[rgb]{0.96, 0.96, 0.96} 2.42 & \cellcolor[rgb]{0.94, 0.94, 0.94} 2.99 & \cellcolor[rgb]{0.98, 0.98, 0.98} 1.13 & \cellcolor[rgb]{0.98, 0.98, 0.98} 1.28 & \cellcolor[rgb]{0.91, 0.91, 0.91} 4.14 & \cellcolor[rgb]{0.98, 0.98, 0.98} 1.29 & \cellcolor[rgb]{0.96, 0.96, 0.96} 2.01 & \cellcolor[rgb]{0.93, 0.93, 0.93} 3.51\\ 
   2015 LA    & \cellcolor[rgb]{1.00, 1.00, 1.00} 0.25 & \cellcolor[rgb]{1.00, 1.00, 1.00} 0.41 & \cellcolor[rgb]{1.00, 1.00, 1.00} 0.24 & \cellcolor[rgb]{1.00, 1.00, 1.00} 0.24 & \cellcolor[rgb]{1.00, 1.00, 1.00} 0.19 & \cellcolor[rgb]{1.00, 1.00, 1.00} 0.31 & \cellcolor[rgb]{1.00, 1.00, 1.00} 0.21 & \cellcolor[rgb]{1.00, 1.00, 1.00} 0.17 & \cellcolor[rgb]{1.00, 1.00, 1.00} 0.17 & \cellcolor[rgb]{1.00, 1.00, 1.00} 0.32 & \cellcolor[rgb]{1.00, 1.00, 1.00} 0.32 & \cellcolor[rgb]{1.00, 1.00, 1.00} 0.34 & \cellcolor[rgb]{1.00, 1.00, 1.00} 0.25 & \cellcolor[rgb]{1.00, 1.00, 1.00} 0.16 & \cellcolor[rgb]{1.00, 1.00, 1.00} 0.26 & \cellcolor[rgb]{1.00, 1.00, 1.00} 0.33 & \cellcolor[rgb]{1.00, 1.00, 1.00} 0.30 & \cellcolor[rgb]{1.00, 1.00, 1.00} 0.30\\ 
2021 LA prog. & \cellcolor[rgb]{0.80, 0.80, 0.80} 7.58 & \cellcolor[rgb]{0.84, 0.84, 0.84} 6.38 & \cellcolor[rgb]{0.83, 0.83, 0.83} 6.56 & \cellcolor[rgb]{0.66, 0.66, 0.66} 10.63 & \cellcolor[rgb]{0.73, 0.73, 0.73} 9.19 & \cellcolor[rgb]{0.85, 0.85, 0.85} 6.27 & \cellcolor[rgb]{0.80, 0.80, 0.80} 7.65 & \cellcolor[rgb]{0.81, 0.81, 0.81} 7.16 & \cellcolor[rgb]{0.79, 0.79, 0.79} 7.82 & \cellcolor[rgb]{0.80, 0.80, 0.80} 7.52 & \cellcolor[rgb]{0.80, 0.80, 0.80} 7.64 & \cellcolor[rgb]{0.72, 0.72, 0.72} 9.43 & \cellcolor[rgb]{0.91, 0.91, 0.91} 4.06 & \cellcolor[rgb]{0.79, 0.79, 0.79} 7.70 & \cellcolor[rgb]{0.81, 0.81, 0.81} 7.28 & \cellcolor[rgb]{0.88, 0.88, 0.88} 5.31 & \cellcolor[rgb]{0.86, 0.86, 0.86} 5.79 & \cellcolor[rgb]{0.79, 0.79, 0.79} 7.70\\ 
2021 LA eval. & \cellcolor[rgb]{0.80, 0.80, 0.80} 7.62 & \cellcolor[rgb]{0.81, 0.81, 0.81} 7.26 & \cellcolor[rgb]{0.81, 0.81, 0.81} 7.18 & \cellcolor[rgb]{0.71, 0.71, 0.71} 9.66 & \cellcolor[rgb]{0.78, 0.78, 0.78} 8.11 & \cellcolor[rgb]{0.84, 0.84, 0.84} 6.53 & \cellcolor[rgb]{0.78, 0.78, 0.78} 7.99 & \cellcolor[rgb]{0.80, 0.80, 0.80} 7.42 & \cellcolor[rgb]{0.80, 0.80, 0.80} 7.61 & \cellcolor[rgb]{0.78, 0.78, 0.78} 7.98 & \cellcolor[rgb]{0.78, 0.78, 0.78} 8.03 & \cellcolor[rgb]{0.71, 0.71, 0.71} 9.74 & \cellcolor[rgb]{0.87, 0.87, 0.87} 5.51 & \cellcolor[rgb]{0.80, 0.80, 0.80} 7.57 & \cellcolor[rgb]{0.80, 0.80, 0.80} 7.39 & \cellcolor[rgb]{0.80, 0.80, 0.80} 7.55 & \cellcolor[rgb]{0.85, 0.85, 0.85} 6.10 & \cellcolor[rgb]{0.79, 0.79, 0.79} 7.80\\ 
2021 DF prog. & \cellcolor[rgb]{0.90, 0.90, 0.90} 4.40 & \cellcolor[rgb]{0.90, 0.90, 0.90} 4.33 & \cellcolor[rgb]{0.91, 0.91, 0.91} 4.14 & \cellcolor[rgb]{0.94, 0.94, 0.94} 3.38 & \cellcolor[rgb]{0.92, 0.92, 0.92} 3.75 & \cellcolor[rgb]{0.93, 0.93, 0.93} 3.55 & \cellcolor[rgb]{0.92, 0.92, 0.92} 3.97 & \cellcolor[rgb]{0.91, 0.91, 0.91} 4.23 & \cellcolor[rgb]{0.89, 0.89, 0.89} 4.94 & \cellcolor[rgb]{0.91, 0.91, 0.91} 4.13 & \cellcolor[rgb]{0.91, 0.91, 0.91} 4.23 & \cellcolor[rgb]{0.88, 0.88, 0.88} 5.08 & \cellcolor[rgb]{0.96, 0.96, 0.96} 2.29 & \cellcolor[rgb]{0.96, 0.96, 0.96} 2.46 & \cellcolor[rgb]{0.93, 0.93, 0.93} 3.68 & \cellcolor[rgb]{0.94, 0.94, 0.94} 3.30 & \cellcolor[rgb]{0.93, 0.93, 0.93} 3.47 & \cellcolor[rgb]{0.90, 0.90, 0.90} 4.59\\ 
2021 DF eval. & \cellcolor[rgb]{0.87, 0.87, 0.87} 5.44 & \cellcolor[rgb]{0.83, 0.83, 0.83} 6.68 & \cellcolor[rgb]{0.85, 0.85, 0.85} 6.18 & \cellcolor[rgb]{0.89, 0.89, 0.89} 4.75 & \cellcolor[rgb]{0.88, 0.88, 0.88} 5.23 & \cellcolor[rgb]{0.88, 0.88, 0.88} 4.98 & \cellcolor[rgb]{0.88, 0.88, 0.88} 5.04 & \cellcolor[rgb]{0.85, 0.85, 0.85} 6.10 & \cellcolor[rgb]{0.86, 0.86, 0.86} 5.88 & \cellcolor[rgb]{0.88, 0.88, 0.88} 5.23 & \cellcolor[rgb]{0.87, 0.87, 0.87} 5.56 & \cellcolor[rgb]{0.87, 0.87, 0.87} 5.36 & \cellcolor[rgb]{0.86, 0.86, 0.86} 5.62 & \cellcolor[rgb]{0.85, 0.85, 0.85} 6.07 & \cellcolor[rgb]{0.85, 0.85, 0.85} 6.08 & \cellcolor[rgb]{0.87, 0.87, 0.87} 5.35 & \cellcolor[rgb]{0.86, 0.86, 0.86} 5.87 & \cellcolor[rgb]{0.89, 0.89, 0.89} 4.95\\ 
\bottomrule
\end{tabular}
}
\end{center}
\vspace{-5mm}
\label{app-tab:result2}
\end{table*}

\subsubsection{Use weighted sum of hidden features from Transformer blocks}
We used the last layer's output from the wav2vec 2.0 model as the front-end features for anti-spoofing. However, the last layer's output may not be the best choice for some down-stream tasks \cite[Sec.3.2]{yang21c_interspeech}. It is also possible to extract all the hidden features from the Transformer blocks in the wav2vec2.0 model and do a weighted sum, i.e., $\boldsymbol{a}_{1:N} = \sum_{i=1}^{K}w_i \boldsymbol{z}_{1:N}^{(i)}$, where $\boldsymbol{z}_{1:N}^{(i)}$ is the output from the $i$-th Transformer block and $w_i$ is the corresponding weight.
The weights $w_i$ are learned on the training set for the down-stream task.  

We test the above strategy on both fixed and fine-tuned SSLs (i.e., Wav2vec 2.0). In the former case, the pre-trained SSL model itself is fixed, and $w_i$ is trained together with the rest of the CM on the ASVspoof 2019 LA training set. In the latter case, $w_i$ is trained with the whole CM. The results are listed in Table~\ref{app-tab:result3}.

The \textbf{weighted sum} strategy is not a clear winner. 
On 2019 LA, we observed different trends on the two pre-trained models -- \textbf{weighted sum} improved the EERs for \texttt{W2V-XLSR} but not for \texttt{W2V-Large2}. On 2021 LA, we observed that \textbf{weighted sum} increased the EERs when using \texttt{W2V-Large2}. 

\begin{table*}[h]
\caption{EERs (\%) on different test sets. All the models were trained using the ASVspoof 2019 LA training set. A darker cell color indicates a higher EER value. Columns with \textbf{a title in bold font are new results}.}
\vspace{-5mm}
\begin{center}
\resizebox{0.7\textwidth}{!}
{
\setlength{\tabcolsep}{4pt}
\begin{tabular}{l|ccc|ccc|ccc|ccc}
\toprule
Front end  &    \multicolumn{6}{c|}{\texttt{W2V-XLSR}, fixed}  &    \multicolumn{6}{c|}{\texttt{W2V-Large2}, fixed}  \\
                 &    \multicolumn{3}{c|}{\shortstack{Feat. from last block \\ (EERs from Table~\ref{tab:result}) }} & \multicolumn{3}{c|}{\textbf{weighted sum}} & \multicolumn{3}{c|}{\shortstack{Feat. from last block \\ (EERs from Table~\ref{tab:result}) }} & \multicolumn{3}{c|}{\textbf{weighted sum}} \\
 \midrule
Back end &  \multicolumn{12}{c}{\BNI} \\
\midrule
             &   I   &  II   &  III  &   I   &  II   &  III  &   I   &  II   &  III  &   I   &  II   &  III \\ 
\midrule
   2019 LA    & \cellcolor[rgb]{0.99, 0.99, 0.99} 1.47 & \cellcolor[rgb]{0.98, 0.98, 0.98} 3.45 & \cellcolor[rgb]{0.97, 0.97, 0.97} 3.77 & \cellcolor[rgb]{1.00, 1.00, 1.00} 0.61 & \cellcolor[rgb]{1.00, 1.00, 1.00} 0.77 & \cellcolor[rgb]{1.00, 1.00, 1.00} 0.90 & \cellcolor[rgb]{1.00, 1.00, 1.00} 0.86 & \cellcolor[rgb]{0.99, 0.99, 0.99} 0.99 & \cellcolor[rgb]{0.99, 0.99, 0.99} 2.08 & \cellcolor[rgb]{0.99, 0.99, 0.99} 1.78 & \cellcolor[rgb]{0.98, 0.98, 0.98} 2.71 & \cellcolor[rgb]{0.96, 0.96, 0.96} 4.89\\ 
   2015 LA    & \cellcolor[rgb]{0.97, 0.97, 0.97} 3.97 & \cellcolor[rgb]{0.95, 0.95, 0.95} 6.78 & \cellcolor[rgb]{0.94, 0.94, 0.94} 8.18 & \cellcolor[rgb]{0.99, 0.99, 0.99} 1.37 & \cellcolor[rgb]{0.99, 0.99, 0.99} 1.54 & \cellcolor[rgb]{0.99, 0.99, 0.99} 1.27 & \cellcolor[rgb]{0.99, 0.99, 0.99} 1.39 & \cellcolor[rgb]{0.99, 0.99, 0.99} 1.39 & \cellcolor[rgb]{0.99, 0.99, 0.99} 1.99 & \cellcolor[rgb]{0.99, 0.99, 0.99} 1.13 & \cellcolor[rgb]{0.99, 0.99, 0.99} 1.67 & \cellcolor[rgb]{0.99, 0.99, 0.99} 1.16\\ 
2021 LA prog. & \cellcolor[rgb]{0.92, 0.92, 0.92} 9.85 & \cellcolor[rgb]{0.83, 0.83, 0.83} 17.29 & \cellcolor[rgb]{0.79, 0.79, 0.79} 20.17 & \cellcolor[rgb]{0.89, 0.89, 0.89} 12.10 & \cellcolor[rgb]{0.88, 0.88, 0.88} 13.25 & \cellcolor[rgb]{0.86, 0.86, 0.86} 14.79 & \cellcolor[rgb]{0.90, 0.90, 0.90} 11.40 & \cellcolor[rgb]{0.91, 0.91, 0.91} 10.50 & \cellcolor[rgb]{0.91, 0.91, 0.91} 10.92 & \cellcolor[rgb]{0.87, 0.87, 0.87} 13.86 & \cellcolor[rgb]{0.86, 0.86, 0.86} 14.74 & \cellcolor[rgb]{0.87, 0.87, 0.87} 14.08\\ 
2021 LA eval. & \cellcolor[rgb]{0.91, 0.91, 0.91} 10.97 & \cellcolor[rgb]{0.81, 0.81, 0.81} 18.91 & \cellcolor[rgb]{0.78, 0.78, 0.78} 20.71 & \cellcolor[rgb]{0.89, 0.89, 0.89} 12.37 & \cellcolor[rgb]{0.86, 0.86, 0.86} 15.06 & \cellcolor[rgb]{0.85, 0.85, 0.85} 15.72 & \cellcolor[rgb]{0.88, 0.88, 0.88} 13.19 & \cellcolor[rgb]{0.89, 0.89, 0.89} 12.57 & \cellcolor[rgb]{0.88, 0.88, 0.88} 12.94 & \cellcolor[rgb]{0.83, 0.83, 0.83} 17.01 & \cellcolor[rgb]{0.83, 0.83, 0.83} 17.08 & \cellcolor[rgb]{0.84, 0.84, 0.84} 16.26\\ 
2021 DF prog. & \cellcolor[rgb]{0.98, 0.98, 0.98} 2.67 & \cellcolor[rgb]{0.96, 0.96, 0.96} 5.09 & \cellcolor[rgb]{0.95, 0.95, 0.95} 7.02 & \cellcolor[rgb]{0.99, 0.99, 0.99} 1.72 & \cellcolor[rgb]{0.99, 0.99, 0.99} 1.89 & \cellcolor[rgb]{0.98, 0.98, 0.98} 3.09 & \cellcolor[rgb]{0.99, 0.99, 0.99} 1.86 & \cellcolor[rgb]{0.99, 0.99, 0.99} 2.12 & \cellcolor[rgb]{0.98, 0.98, 0.98} 3.36 & \cellcolor[rgb]{0.98, 0.98, 0.98} 3.55 & \cellcolor[rgb]{0.96, 0.96, 0.96} 5.39 & \cellcolor[rgb]{0.95, 0.95, 0.95} 7.32\\ 
2021 DF eval. & \cellcolor[rgb]{0.95, 0.95, 0.95} 7.14 & \cellcolor[rgb]{0.92, 0.92, 0.92} 9.94 & \cellcolor[rgb]{0.90, 0.90, 0.90} 11.35 & \cellcolor[rgb]{0.93, 0.93, 0.93} 9.11 & \cellcolor[rgb]{0.94, 0.94, 0.94} 7.72 & \cellcolor[rgb]{0.93, 0.93, 0.93} 8.78 & \cellcolor[rgb]{0.95, 0.95, 0.95} 7.44 & \cellcolor[rgb]{0.94, 0.94, 0.94} 7.77 & \cellcolor[rgb]{0.93, 0.93, 0.93} 9.26 & \cellcolor[rgb]{0.93, 0.93, 0.93} 8.91 & \cellcolor[rgb]{0.92, 0.92, 0.92} 9.99 & \cellcolor[rgb]{0.91, 0.91, 0.91} 10.57\\ 
\bottomrule
 \multicolumn{13}{c}{}\\
\toprule
Front end  &    \multicolumn{6}{c|}{\texttt{W2V-XLSR}, fine-tuned}  &    \multicolumn{6}{c|}{\texttt{W2V-Large2}, fine-tuned}  \\
                 &    \multicolumn{3}{c|}{\shortstack{Feat. from last block \\ (EERs from Table~\ref{tab:result}) }} & \multicolumn{3}{c|}{\textbf{weighted sum}} & \multicolumn{3}{c|}{\shortstack{Feat. from last block \\ (EERs from Table~\ref{app-tab:result}) }} & \multicolumn{3}{c|}{\textbf{weighted sum}} \\
 \midrule
Back end &  \multicolumn{12}{c}{\BNI} \\
\midrule
             &   I   &  II   &  III  &   I   &  II   &  III  &   I   &  II   &  III  &   I   &  II   &  III \\ 
\midrule
  2019 LA    & \cellcolor[rgb]{0.99, 0.99, 0.99} 2.31 & \cellcolor[rgb]{0.98, 0.98, 0.98} 2.80 & \cellcolor[rgb]{0.98, 0.98, 0.98} 3.08 & \cellcolor[rgb]{1.00, 1.00, 1.00} 0.44 & \cellcolor[rgb]{1.00, 1.00, 1.00} 0.76 & \cellcolor[rgb]{0.98, 0.98, 0.98} 3.53 & \cellcolor[rgb]{0.99, 0.99, 0.99} 1.86 & \cellcolor[rgb]{0.98, 0.98, 0.98} 2.42 & \cellcolor[rgb]{0.98, 0.98, 0.98} 2.99 & \cellcolor[rgb]{0.97, 0.97, 0.97} 3.81 & \cellcolor[rgb]{0.96, 0.96, 0.96} 6.11 & \cellcolor[rgb]{0.93, 0.93, 0.93} 9.20\\ 
   2015 LA    & \cellcolor[rgb]{1.00, 1.00, 1.00} 0.25 & \cellcolor[rgb]{1.00, 1.00, 1.00} 0.41 & \cellcolor[rgb]{1.00, 1.00, 1.00} 0.24 & \cellcolor[rgb]{1.00, 1.00, 1.00} 0.32 & \cellcolor[rgb]{1.00, 1.00, 1.00} 0.51 & \cellcolor[rgb]{1.00, 1.00, 1.00} 0.33 & \cellcolor[rgb]{1.00, 1.00, 1.00} 0.32 & \cellcolor[rgb]{1.00, 1.00, 1.00} 0.32 & \cellcolor[rgb]{1.00, 1.00, 1.00} 0.34 & \cellcolor[rgb]{1.00, 1.00, 1.00} 0.62 & \cellcolor[rgb]{1.00, 1.00, 1.00} 0.48 & \cellcolor[rgb]{0.99, 0.99, 0.99} 1.26\\ 
2021 LA prog. & \cellcolor[rgb]{0.94, 0.94, 0.94} 7.58 & \cellcolor[rgb]{0.95, 0.95, 0.95} 6.38 & \cellcolor[rgb]{0.95, 0.95, 0.95} 6.56 & \cellcolor[rgb]{0.96, 0.96, 0.96} 5.13 & \cellcolor[rgb]{0.97, 0.97, 0.97} 4.42 & \cellcolor[rgb]{0.96, 0.96, 0.96} 5.01 & \cellcolor[rgb]{0.94, 0.94, 0.94} 7.52 & \cellcolor[rgb]{0.94, 0.94, 0.94} 7.64 & \cellcolor[rgb]{0.92, 0.92, 0.92} 9.43 & \cellcolor[rgb]{0.89, 0.89, 0.89} 12.06 & \cellcolor[rgb]{0.92, 0.92, 0.92} 9.65 & \cellcolor[rgb]{0.86, 0.86, 0.86} 15.21\\ 
2021 LA eval. & \cellcolor[rgb]{0.94, 0.94, 0.94} 7.62 & \cellcolor[rgb]{0.95, 0.95, 0.95} 7.26 & \cellcolor[rgb]{0.95, 0.95, 0.95} 7.18 & \cellcolor[rgb]{0.96, 0.96, 0.96} 4.97 & \cellcolor[rgb]{0.96, 0.96, 0.96} 5.34 & \cellcolor[rgb]{0.96, 0.96, 0.96} 5.62 & \cellcolor[rgb]{0.94, 0.94, 0.94} 7.98 & \cellcolor[rgb]{0.94, 0.94, 0.94} 8.03 & \cellcolor[rgb]{0.92, 0.92, 0.92} 9.74 & \cellcolor[rgb]{0.88, 0.88, 0.88} 12.99 & \cellcolor[rgb]{0.91, 0.91, 0.91} 10.44 & \cellcolor[rgb]{0.85, 0.85, 0.85} 15.75\\ 
2021 DF prog. & \cellcolor[rgb]{0.97, 0.97, 0.97} 4.40 & \cellcolor[rgb]{0.97, 0.97, 0.97} 4.33 & \cellcolor[rgb]{0.97, 0.97, 0.97} 4.14 & \cellcolor[rgb]{1.00, 1.00, 1.00} 0.88 & \cellcolor[rgb]{0.99, 0.99, 0.99} 1.42 & \cellcolor[rgb]{0.99, 0.99, 0.99} 2.10 & \cellcolor[rgb]{0.97, 0.97, 0.97} 4.13 & \cellcolor[rgb]{0.97, 0.97, 0.97} 4.23 & \cellcolor[rgb]{0.96, 0.96, 0.96} 5.08 & \cellcolor[rgb]{0.96, 0.96, 0.96} 6.15 & \cellcolor[rgb]{0.95, 0.95, 0.95} 6.66 & \cellcolor[rgb]{0.90, 0.90, 0.90} 11.11\\ 
2021 DF eval. & \cellcolor[rgb]{0.96, 0.96, 0.96} 5.44 & \cellcolor[rgb]{0.95, 0.95, 0.95} 6.68 & \cellcolor[rgb]{0.96, 0.96, 0.96} 6.18 & \cellcolor[rgb]{0.96, 0.96, 0.96} 5.18 & \cellcolor[rgb]{0.96, 0.96, 0.96} 6.16 & \cellcolor[rgb]{0.96, 0.96, 0.96} 5.00 & \cellcolor[rgb]{0.96, 0.96, 0.96} 5.23 & \cellcolor[rgb]{0.96, 0.96, 0.96} 5.56 & \cellcolor[rgb]{0.96, 0.96, 0.96} 5.36 & \cellcolor[rgb]{0.95, 0.95, 0.95} 6.36 & \cellcolor[rgb]{0.96, 0.96, 0.96} 6.19 & \cellcolor[rgb]{0.93, 0.93, 0.93} 8.79\\ 
\bottomrule
\end{tabular}
}
\end{center}
\vspace{-5mm}
\label{app-tab:result3}
\end{table*}

\end{document}